\documentclass[fleqn,usenatbib]{mnras}

\usepackage{newtxtext,newtxmath}
\usepackage[T1]{fontenc}
\usepackage[normalem]{ulem}

\DeclareRobustCommand{\VAN}[3]{#2}
\let\VANthebibliography\thebibliography
\def\thebibliography{\DeclareRobustCommand{\VAN}[3]{##3}\VANthebibliography}

\usepackage{graphicx}	% Including figure files
\usepackage{amsmath}	% Advanced maths commands
\usepackage{subcaption}
\usepackage{float}
\usepackage{xcolor}
\usepackage{multirow}
\usepackage{orcidlink}
\newcommand*{\mailto}[1]{\href{mailto:#1}{#1}}

\newcommand{\msun}{M$_{\odot}$}

\title[ICL and intra-cluster GCs in TNG50]{The progenitors of the intra-cluster light and intra-cluster globular clusters in galaxy groups and clusters}

\author[N. Ahvazi et al.]{Niusha Ahvazi,$^{\orcidlink{0009-0002-1233-2013}1,2}$\thanks{E-mail:  \mailto{niusha.ahvazi@email.ucr.edu} / \mailto{nahvazi@carnegiescience.edu}}
Laura V. Sales,$^{\orcidlink{0000-0002-3790-720X}1}$
Jessica E. Doppel$^{\orcidlink{0000-0001-5354-4229}1}$
Andrew Benson,$^{\orcidlink{0000-0001-5501-6008}2}$
Richard D'Souza$^{\orcidlink{0000-0001-9269-8167}3}$ \and
 and Vicente Rodriguez-Gomez$^{\orcidlink{0000-0002-9495-0079}4}$
\\\\
$^{1}$Department of Physics and Astronomy, University of California, Riverside, 900 University Avenue, Riverside, CA 92521, USA\\
$^{2}$Carnegie Observatories, 813 Santa Barbara Street, Pasadena, CA 91101, USA\\
$^{3}$Vatican Observatory, Specola Vaticana, V-00120, Vatican City State\\
$^{4}$Instituto de Radioastronom\'ia y Astrof\'isica, Universidad Nacional Aut\'onoma de M\'exico, Apdo. Postal 72-3, 58089 Morelia, Mexico
}
\vspace{0.2cm}

\date{Accepted XXX. Received YYY; in original form ZZZ}

\pubyear{2023}

\begin{document}
\label{firstpage}
\pagerange{\pageref{firstpage}--\pageref{lastpage}}
\maketitle

\begin{abstract}
We use the IllustrisTNG50 cosmological hydrodynamical simulation, complemented by a catalog of tagged globular clusters, to investigate the properties and build up of two extended luminous components: the intra-cluster light (ICL) and the intra-cluster globular clusters (ICGC). We select the 39 most massive groups and clusters in the box, spanning the range of virial masses $5 \times 10^{12} < \rm M_{200}/\rm M_{\odot} < 2 \times 10^{14}$. We find good agreement between predictions from the simulations and current observational estimates of the fraction of mass in the ICL and its radial extension. The stellar mass of the ICL is only $\sim10\%$--$20\%$ of the stellar mass in the central galaxy but encodes useful information on the assembly history of the group or cluster. About half the ICL in all our systems is brought in by galaxies in a narrow stellar mass range, $M_*=10^{10}$--$10^{11}$\msun. However, the contribution of low-mass galaxies ($M_*<10^{10}$\msun) to the build-up of the ICL varies broadly from system to system, $\sim 5\%-45\%$, a feature that might be recovered from the observable properties of the ICL at $z=0$. At fixed virial mass, systems where the accretion of dwarf galaxies plays an important role have shallower metallicity profiles, less metal content and a lower stellar mass in the ICL than systems where the main contributors are more massive galaxies. We show that intra-cluster GCs are also good tracers of this history, representing a valuable alternative when diffuse light is not detectable. 
\end{abstract}


%%%%%%%%%%%%%%%%%%%%%%%%%%%%%%%%%%%%%%%%%%%%%%%%%%

%%%%%%%%%%%%%%%%% BODY OF PAPER %%%%%%%%%%%%%%%%%%

\section{Introduction}

One of the most striking features of high density environments is the intra-cluster light (ICL), a diffuse component of light that originates from populations of stars that are not associated with individual galaxies and are instead gravitationally bound to the host dark matter halo. The ICL is thought to be a product of the tidal stripping of stars from galaxies as they traverse the cluster or group environments and was first proposed and subsequently discovered by \cite{Zwicky1, Zwicky2, Zwicky3, Zwicky4} in the Coma cluster. The direct link between the ICL and satellite galaxies makes the formation of this diffuse light a natural prediction of the hierarchical assembly model in Cold Dark Matter \citep[CDM, ][]{White1978, Conroy2007, Montes2018, Contini2021}.

The ICL has cosmological relevance primarily in two aspects. First, it represents a visible tracer of the unseen dark matter distribution, with several theoretical works supporting a good correlation between the shape and orientation of the ICL and those of the underlying dark matter halo \citep{Montes2019, Alonso2020, Contini2020, Deason2020, Gonzalez2021}. Second, and the main focus of this paper, the ICL is built by the tidal disruption of many satellite galaxies, some of which do not survive until today. As such it can help unravel the past formation history of the host group or cluster halo (\citealt{MontesReview2022, Contini2021}, and references therein), in a similar way that stellar halos can help reconstruct the merger histories of smaller mass halos in the MW-like regime \citep[e.g.,][]{Bullock2005}.

%mass content of ICL
The stellar mass content of the ICL is directly related to the stellar mass-halo mass relation and serves as a probe of the past assembly history of galaxy clusters. A number of observational constraints on the amount of ICL have been reported in the literature, with studies using deep imaging to estimate the total amount of ICL in galaxy clusters (e.g. \citealt{Zibetti2005, Mihos2017, Krick2007, Morishita2017, Jimenez2019, McGee2010}). These studies have found that the ICL typically makes up a significant fraction of the total light in galaxy clusters, ranging from <10\% to 50\% depending on the cluster and the methods used to estimate the ICL \citep[see][]{MontesReview2022}.

Numerical simulations predict that the bulk of the ICL mass comes from the tidal stripping of massive satellites (10 < log(M/M$_{\odot}$) < 11) \citep{Puchwein2010, Cui2014, Cooper2015, Contini2013, Contini2019, Montenegro2023}. Particularly, disk-like massive satellites are thought to significantly contribute to building the ICL through a large number of small stripping events \citep{Contini2018}. However, other less-dominant mechanisms may also contribute stars to the ICL including the total disruption of low mass satellites \citep{Purcell2007}, stars ejected into the inter-cluster medium after a major merger \citep{Murante2007} and the pre-processing of accreted groups \citep{C.Rudick2006}. Due to the well established relation between stellar mass and metallicity in galaxies, the nature of the progenitors that build up the ICL can be observationally constrained from stellar metallicities. Furthermore, each of these mechanisms are expected to leave distinct patterns on the metallicity and their gradients.

Constraining the contribution of each mechanism can be extremely challenging due to the faint ($\mu_{V} \sim 26.5\; \rm mag/arcsec^{2}$) and extended characteristics of the ICL, which can often only be probed by broad-band photometry. Yet, over the last two decades, significant progress has been made. A number of observational studies have highlighted the presence of clear negative radial color gradients \citep{Krick2007, Rudick2009, Melnick2012, DeMaio2015} in the ICL of the majority of clusters studied at $z \sim 0.5$. While such gradients could potentially arise due to changes in metallicity \citep{Montes2014, DeMaio2015} or variations in the ages of the stars \citep{Morishita2017, Montes2018}, it suggests that violent relaxation after major mergers with the BCG cannot be the dominant source of ICL. Although, the observed metallicities of the ICL ($\mathrm{[Fe/H]_{ICL}} \sim -0.5$) align with the notion that the ICL stars likely originate either from stars located in the outer regions of galaxies with stellar masses approximately $5 \times 10^{10}$M$_{\odot}$ \citep{DeMaio2018,Montes2018, Montes2021} or from the dissolution of dwarf galaxies, the amount of ICL is often used as an argument against the latter as this would dramatically alter the faint end of the cluster galaxy luminosity function \citep[e.g.][]{Zibetti2005, DeMaio2018}. Only a small number of clusters have been identified with flat color gradients (e.g. Abell 370) - indicating an ICL formed through the expulsion of stars into the intra-cluster medium during a major merger event \citep{Krick2007, DeMaio2015, DeMaio2018}. While the vast variety of observations point to the progenitors of the ICL being the tidal stripping of  massive satellites (10 < log(M/M$_{\odot}$) < 11) consistent with theoretical predictions, the ICL of the two closest well-studied clusters, the Virgo and Coma, which can be studied in significant detail, suggest very different assembly mechanisms. \cite{Williams2007}, studying the RGB populations of the Virgo cluster ($\sim$15 Mpc) using a single and deep HST pointing, found that 70\% of the stars have a metallicity of $\mathrm{[M/H]} \sim -1.3$) indicating that the ICL was built up primarily through the disruption of dwarf galaxies. Similarly, \cite{Gu2020} spectroscopically studied the ICL of the Coma cluster in the low S/N regime finding it to be old and metal-poor ($\mathrm{[M/H]} \sim -1.0$) -- consistent with the accretion of low-mass galaxies or the tidal stripping of the outskirts of massive galaxies that have ended their star formation early on. The low metallicity of the ICL of the Virgo and the Coma cluster clearly suggests that the contribution of dwarf galaxies to the ICL can be significant.  This is clearly in tension with the observations of clusters further away as well as the current theoretical paradigm, requiring further study and perhaps a revision of understanding of the contribution of dwarf galaxies to the ICL.

One way to observationally disentangle the various formation mechanism is to study how the ICL correlates with the mass of the cluster and with redshift. However, as we probe deeper into the universe and examine systems at earlier redshifts, the low surface brightness of the intra-cluster light becomes increasingly difficult to capture. In such cases, an alternative path to study the dark matter halos may be offered by globular clusters (GCs). These ancient, dense clusters of stars are commonly found in nearly all types of galaxies and are thought to be among the oldest and (often) metal-poor stellar populations in the universe \citep{Gratton_2019}. They are believed to have formed before the majority of galaxies and are relatively luminous \citep{Schauer2021}, making them easily detectable at large distances and earlier redshifts and a powerful tool to study the distribution of dark matter and ICL in the early universe. 

GCs can be very numerous in groups and clusters. While for MW-like galaxies they populate host halos by the hundreds, in the Virgo cluster more than 10,000 have been cataloged around M87 \citep{Durrell2014}. Studies have shown that the abundance of GCs is closely related to the amount of dark matter in the halos \citep{Spitler2009, Harris_Catalog, Hudson1, Harris2015}. Therefore, GCs can offer an alternative path to study the dark matter halos in galaxy groups and clusters, especially when the ICL is not easily captured. 

A large fraction of GCs associated with a given system are members of the intra-cluster globular cluster (ICGC) component, which has been observationally confirmed in Fornax \citep{Schuberth_2007}, Coma \citep{Madrid2018}, Abell 1689 \citep{Alamo2017}, Virgo \citep{Lee2010, Ko2017, Longobardi2018} and Centaurus A \citep{Taylor2017} for example. This intra-cluster component is expected to arise mostly by the gravitational removal of GCs from satellite galaxies that interacted with the groups, and predominantly from satellites that did not survive until today \citep{Ramos-Almendares2020}. In this way, the origin of the ICGCs and the ICL are therefore expected to be strongly linked, and one might naively expect similar properties and building blocks for both components. However, the way stars and GCs occupy halos is different, in particular in the low mass end of dwarf galaxies, and it is as yet unclear how this may impact the way GCs and the ICL trace each other. 

In this study, we take advantage of the recently published catalog of GCs\footnote{\url{www.tng-project.org/doppel22}}, which are tagged post-processing into the 
cosmological hydrodynamical simulation IllustrisTNG50 to explore the properties of the ICL and GC components of groups and clusters. The paper is organized as follows: in section~\ref{sec:simulation}, we present a brief description of the simulation and definitions used in our study. In section~\ref{sec:properties}, we introduce the general properties of the ICL component predicted by the simulation. In section~\ref{Prog}, we analyze the progenitors of the ICL component, and in section~\ref{GCs}, we investigate the use of GCs as tracers of the formation history of ICL. Finally, in section~\ref{sec:conclusion}, we summarize the main results of our study.

\section{Simulation}\label{sec:simulation}

We use the cosmological hydrodynamical simulation Illustris-TNG50 \citep[TNG50 for short, ][]{Pillepich2019,Nelson2019TNG50}, which is part of the suite of cosmological boxes from the  Illustris-TNG project\footnote{\url{https://www.tng-project.org}} \citep{Pillepich2018, Springel2018, Marinacci2018, Naiman2018, Nelson2018, Nelson2019}. TNG50 is the largest resolution baryonic run of the Illustris-TNG suite, which has a volume of approximately $50 ^{3} \, \rm Mpc^{3}$ and $2 \times 2160^3$ resolution elements, with an average mass per particle $8.5 \times 10^4 \rm M_{\odot}$ and $4.5 \times 10^5 \rm M_{\odot}$ for baryons and dark matter, respectively. The Illustris-TNG cosmological parameters are consistent with $\Lambda$CDM model determined by Planck XIII \citep{Planck_Collaboration2016} to be $\Omega_\mathrm{m} = 0.3089,\, \Omega_\mathrm{b} = 0.0486,\, \Omega_{\Lambda} = 0.6911,\, H_{0} = 100 \, h \; \rm km \; \rm s^{-1} \; \rm Mpc^{-1}$ with $h = 0.6774,\, \sigma_{8} = 0.8159,$ and $n_\mathrm{s} = 0.9667.$

The evolution of gravity and hydrodynamics are followed using the {\sc arepo} moving mesh code \citep{Springel2010}. The galaxy formation baryonic treatment is based on its predecessor simulation suite, Illustris \citep{Vogelsberger2013, Vogelsberger2014, Vogelsberger2014Nature, Genel2014, Nelson2015}, with modifications implemented to better track the formation and evolution of galaxies, as described in \cite{Pillepich2018MNRAS, Weinberger2017}. 

The updated Illustris-TNG sub-grid models accounts for star formation, radiative metal cooling, chemical enrichment from SNII, SNIa, and AGB stars, stellar feedback, and super-massive black hole feedback \citep{Weinberger2017, Pillepich2018MNRAS}. These models are shown to reproduce several of the $z=0$ basic galaxy scaling relations, including the stellar mass - size \citep{Genel2018}, the color bimodality \citep{Nelson2018} and galaxy clustering \citep{Springel2018}, among others. As such, they are representative of the present day population of galaxies in the universe and reproduce the main environmental trends observed in satellites. Of particular relevance to this work, the abundance of low-mass and intermediate mass galaxies seems consistent with observationally estimated stellar mass functions \citep[e.g.][]{Pillepich2018MNRAS,Vazquez-Mata2020, Engler2021}.

\subsection{Identification of groups and clusters}

Groups are identified using spatial information based on a Friends-of-Friends (FoF) algorithm \citep{Davis1985}. Individual self-gravitating subhalos and galaxies are later found in these groups using  {\verb'SUBFIND'} \citep{Springel2001, Dolag2009}. The object at the center of the gravitational potential of each group is called the “central” galaxy, while all other substructures are “satellites” (or "subhalos"). {\verb'SUBFIND'} identifies substructures with a minimum of $32$ particles and we additionally apply $\rm M_{\rm DM}\geqslant 5.4 \times10^{7}\; \rm M_\odot$ (at the time of infall) to remove the chances of including spurious baryonic clumps that are not bonafide galaxies. The time evolution of galaxies and halos through the 99 snapshots of the simulation is followed by using the SubLink merger trees \citep{Rodriguez2015}.

We identify groups and low mass clusters in TNG50 by selecting all host halos with virial masses $M_{200}/\rm M_{\odot} \geqslant 5 \times 10^{12}$, where virial quantities are measured within the virial radius, or $r_{200}$, defined as the radius of a sphere where the mean density of the group is $200$ times the critical density of the Universe. This selection results in $39$ groups and clusters, containing $3305$ and $5020$ satellite galaxies with $M_*> 10^7 \; \rm M_\odot$ within the virial radius and FoF group boundaries, respectively.

%%%%%%%%%%%%%%%%%%%%%%%%%%%%%%%%%%%%%%%%%%%%%%%%
\begin{figure*}
	\includegraphics[width=\columnwidth]{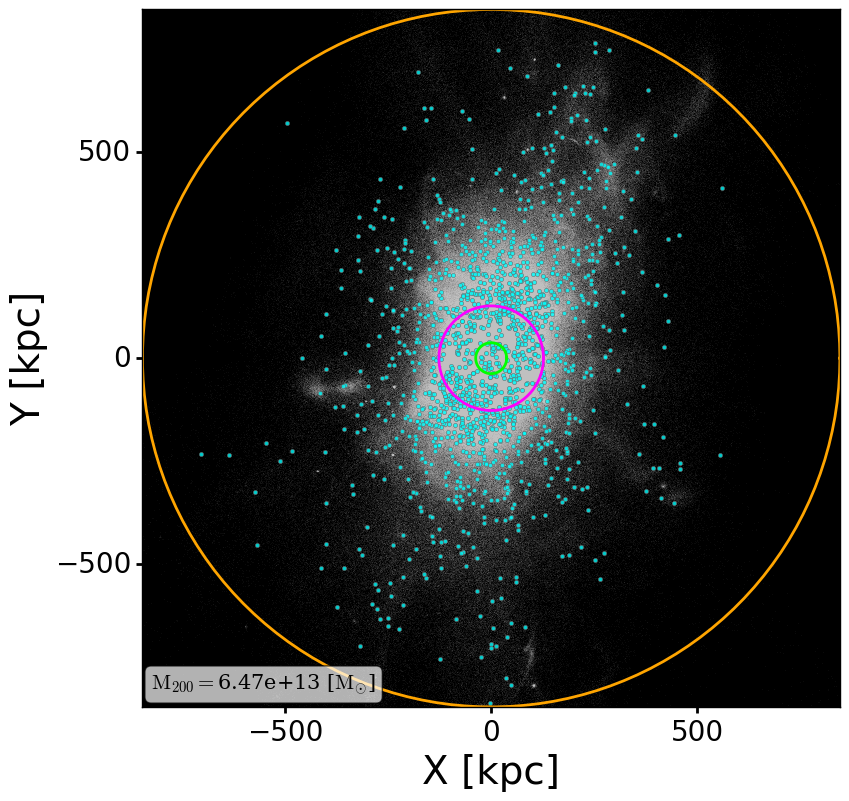}
\includegraphics[width=\columnwidth]{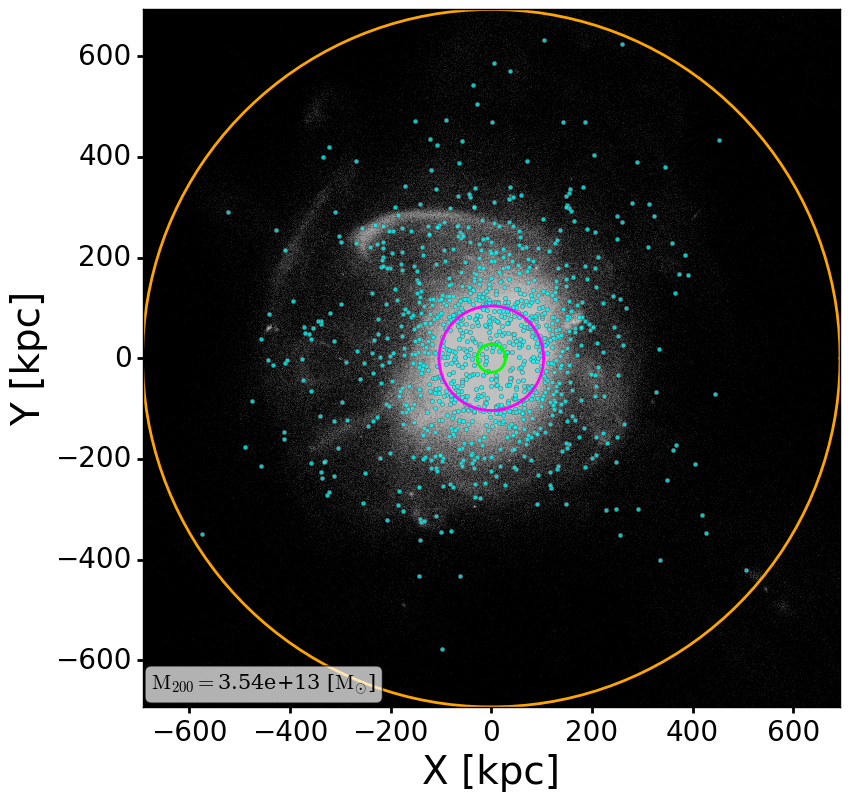}

    \includegraphics[width=\columnwidth]{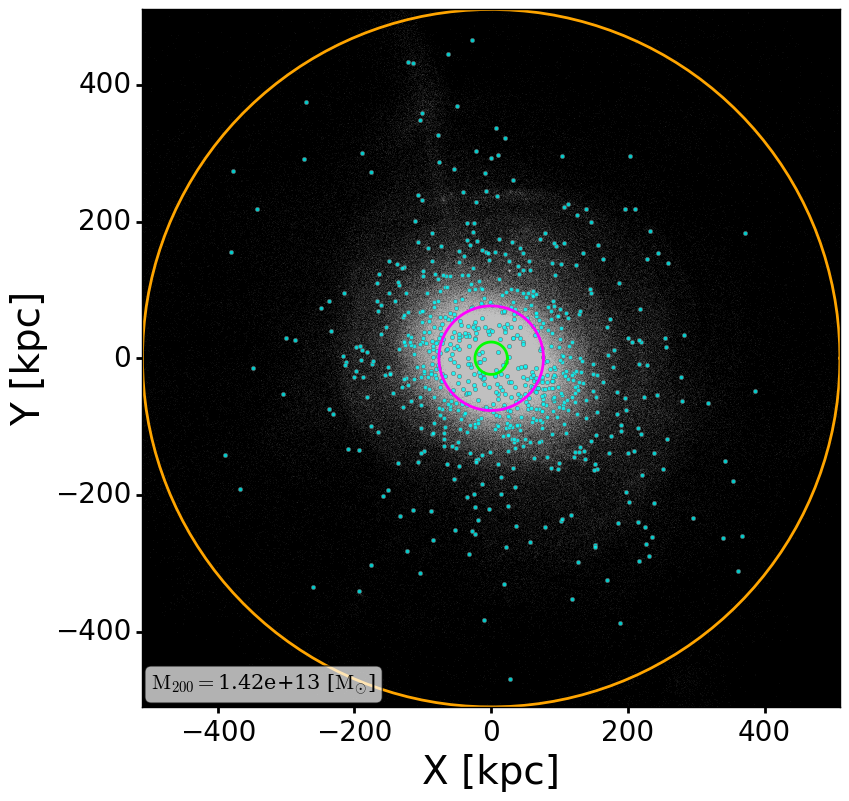}
\includegraphics[width=\columnwidth]{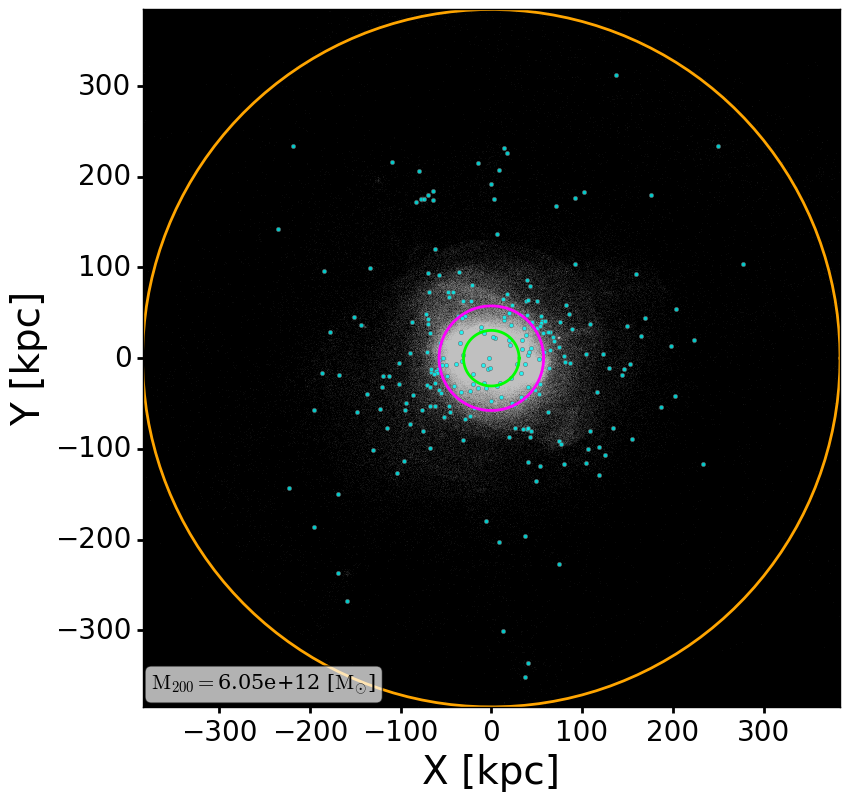}

    \caption{Projected position of the luminous (stellar (gray-scale background) + ICGCs (cyan dots)) components in the ICL of 4 randomly selected groups from our sample (the virial mass of each group is quoted on the bottom left of each panel). Orange, magenta, and lime circles show the position of $\rm r_{200}$ for the group, and $0.15 \times \rm r_{200}$, and $2 \times \rm r_{h_{*}}$, respectively.}
    \label{fig1:proj. pos.}
\end{figure*}
%%%%%%%%%%%%%%%%%%%%%%%%%%%%%%%%%%%%%%%%%%%%%%%%

\subsection{The intra-cluster light (ICL) in groups and clusters}
\label{method2}

For the purpose of this work we define the ICL as all stellar particles that are unbound to any substructure, (exclusively) bound to the group as a whole, and located within a radial range of $0.15 \, \rm r_{200} < \rm r < \rm r_{200}$ from the host galaxy, where $\rm r_{200}$ is the virial radius of the group. Different definitions of ICL are commonly assumed in the literature, both in simulations and observational work. We have experimented with several of these definitions and explicitly checked that none of our main conclusions depends qualitatively on the specific criteria adopted here. For a more detailed view, we collect in  Appendix~\ref{App A}  some examples on how the exact definition of ICL affects some of our reported results.

Further inspection of our selection of ICL particles revealed an excess of stars located on the periphery of several satellites. These stellar particles are considered unbound by {\verb'SUBFIND'} (and therefore included as ICL candidates) but are still clearly part of the satellite or substructure in the 6D space of positions and velocities. We therefore applied two extra requirements in order to ``clean" our ICL sample. Each stellar particle needs to be at least $10 \times \rm r_{h_{*}}$ away from any massive satellite (the ones larger than $\rm M/\rm M_{\odot} > 10^{10}$), where $\rm r_{h_{*}}$ is the stellar half mass radius of the satellite. In addition, we ensure that the velocity of the particles satisfy V/$\rm V_{C}$ $>$ 2.5 to be considered part of the ICL, where $\rm V_{C}$ is the circular velocity at $2 \times \rm r_{h_{*}}$ of the satellite. This procedure satisfactorily removes extra stellar particles with positions and velocities strongly correlated with the local substructure. Note that similar methods are used in observations of the ICL to reduce the light contamination from the central galaxy and the satellite galaxies. 

\subsection{In-situ vs. ex-situ}
\label{ssec:insitu}

In this paper, we frequently use the terms ``in-situ" and ``ex-situ" to distinguish between different types of stellar populations: those born from gas bound to the main branch progenitor in the SubLink merger tree of a galaxy (``in-situ'') versus those born in external galaxies and later brought in to the descendant object during mergers or tidal stripping events (``ex-situ''). 

To classify stars as ex-situ or in-situ, we use the stellar assembly catalogs, offered as auxiliary data in the TNG database. The classification method was introduced in detail in \citet{Rodriguez2016} and is based on the {\verb'SUBFIND'} association of the newly formed stars (at the time of birth) to either the main progenitor of the central galaxy (in-situ) or to a substructure that later merges into the system (ex-situ). Main or secondary progenitors are defined using the SubLink merger trees.

\subsection{Globular Clusters}\label{method4}

%%%%%%%%%%%%%%%%
\begin{figure*}
	\includegraphics[width=\columnwidth]{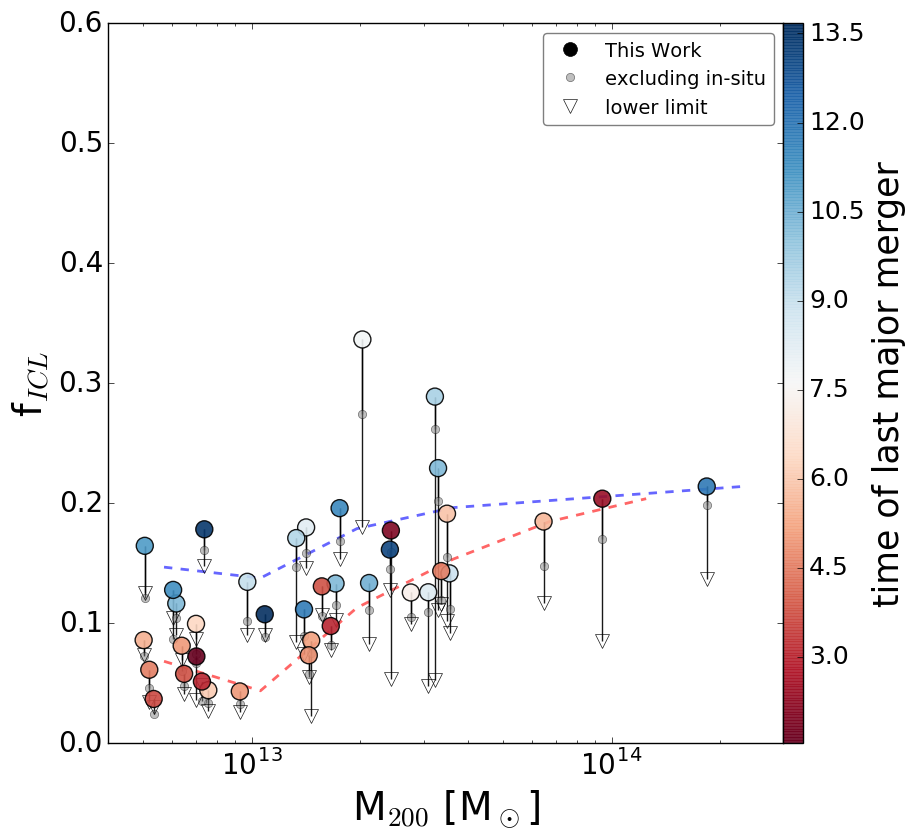}
	\includegraphics[width=\columnwidth]{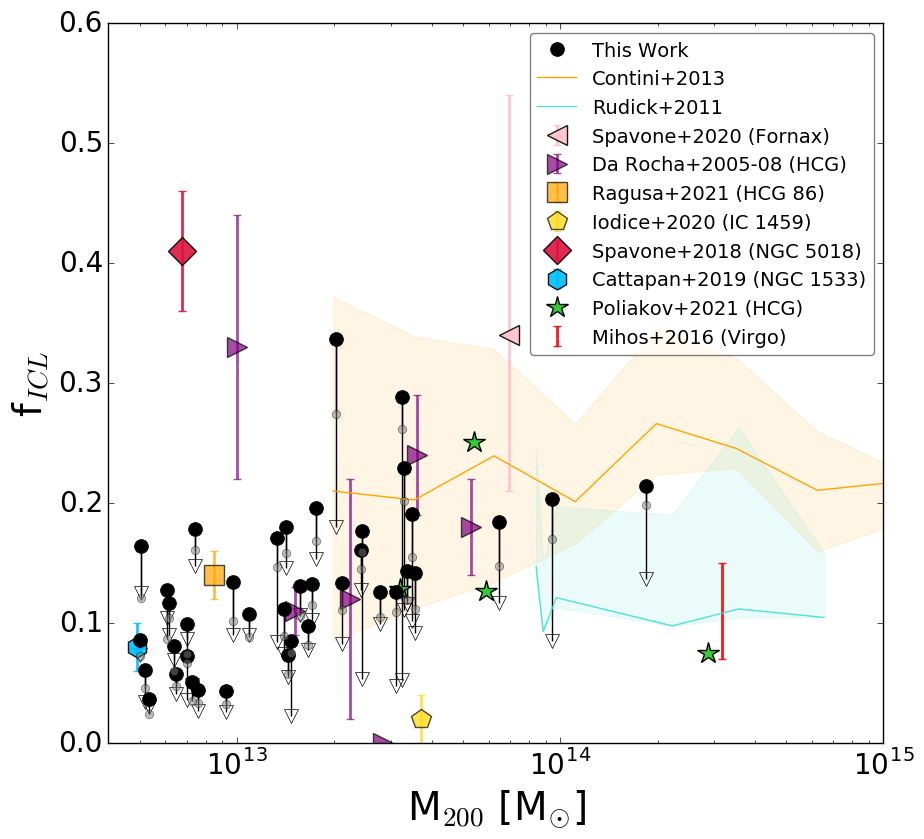}
    \caption{ICL fraction vs virial mass of the halos. Predictions by TNG50 are shown with filled large circles (upper limit), and empty downward triangles (lower limit). Fraction of only ex-situ stellar component in ICL are shown by grey circles. Left panel: ICL fractions are colored by the time of the last major merger occurred for the central galaxy. The light 'red' and 'blue' dotted lines show the average of ICL fraction for the corresponding red and blue points. Right panel: Results from other simulation are shown by the shaded regions, and several observational studies are shown with different colored markers for comparison. references along with the reported quantities are presented in table~\ref{tab:obs_data} in Appendix~\ref{app:obs}.}
    \label{fig2:f_ICL}
\end{figure*}
%%%%%%%%%%%%%%%%

We use the catalog of GCs presented in \citet{GCCat} for our TNG50 systems. GCs are tagged for all galaxies with $\rm M_{*, max} \geq 5\times 10^{6}$ M$_{\odot}$ that infall or interact with our groups and clusters. The tagging is done at infall time for all galaxies, after which the dynamics of the GCs is followed naturally by tracking the positions and velocities of the particles flagged as GC candidates. The simulated GCs are shown to follow known scaling relations for galaxy-GCs and, in addition, to give rise to a population of intra-cluster GCs (ICGCs) as a combination of tidal stripping of GCs from merging galaxies and a native GC population tagged onto the central galaxy \citep{Ramos-Almendares2020,GCCat}. The slope and normalization of this ICGC component is in rough agreement with current observational constraints \citep[see Fig. 5 in ][]{GCCat}. In Appendix~\ref{app:gcs} we summarize the technical details of the GC tagging technique, for additional information we refer the reader to \citet{GCCat}.

For this work, we use from this catalog the ICGCs, identified as tagged GCs that are currently not associated to any substructure according to {\verb'SUBFIND'} and ``cleaned" with the same method described in section~\ref{method2} for the ICL. As an example of how these GCs are distributed in our systems, we show in Fig.~\ref{fig1:proj. pos.}, projected maps of multiple groups from our sample. These plots show the stellar component (gray-scale in the background) with the position of the ICGCs highlighted by the cyan dots. These few examples illustrate interesting differences in size, shapes and concentration of the GC and ICL components, suggesting a link to their different formation histories.  

\section{The mass and extent predicted for the ICL} \label{sec:properties}

\subsection{ICL mass content}
\label{amount}

In order to quantify the amount of mass in the ICL we define the fraction of mass in the ICL as: 

\begin{equation}
 f_\mathrm{ICL}=\frac{M_{ICL}^{*}}{M_{BCG}^{*}},
 \label{eq:icl}
\end{equation}

where $\rm M_{ICL}^{*}$ is the stellar mass in the ICL and $\rm M_{BCG}^{*}$ is the stellar mass within $2 \times \rm r_{h_{*}}$ of the central galaxy. Solid circles with black outlines in Fig~\ref{fig2:f_ICL} show $f_\mathrm{ICL}$ for our sample of groups and clusters plotted against the virial mass of each group. We note that various definitions of this fraction have been used in the literature. In particular, a lower limit to $f_{\rm ICL}$ might be found by dividing the mass of the ICL by the total mass of stars within the virial radius $f_{\rm ICL,tot}=M_{\rm ICL}/M_{*,r_{200}}$. For comparison, we show $f_{\rm ICL,tot}$ in Fig.~\ref{fig2:f_ICL} with downward triangles connected to the solid symbols. In general we find that the ICL fraction can change by more than a factor of two by adopting common definitions in the literature (see Appendix~\ref{App A}). In this work, $f_{\rm ICL}$ will refer to Eq.~\ref{eq:icl} unless specified otherwise.

Our systems span from $5 \times 10^{12}$ to nearly $2 \times 10^{14}$ $\rm M_{\odot}$ in virial mass, sampling the range from isolated elliptical galaxies comparable to Centaurus A to moderate mass galaxy clusters such as Fornax, and conservative mass estimates of the Virgo cluster. The ICL mass fraction shows a subtle increase with virial mass, with typical values $5$-$10$\% in our low mass systems and $\sim 20\%$ for our more massive clusters. The relation also exhibits a substantial degree of dispersion at a given mass.

We trace back the dispersion in this relation to the assembly history of each group. The color coding in the left panel of Fig.~\ref{fig2:f_ICL} shows that, at fixed halo mass, groups with a longer time since their last major merger tend to show lower ICL mass fractions than groups where the last major merger was more recent. This trend has been quantified by categorizing the systems into ``red" and ``blue" groups, based on whether their last major merger occurred within 7.5 Gyr or after. Major mergers are here defined as merger events of the central galaxy with satellites of stellar mass ratios greater than $0.25$ measured at the time of maximum stellar mass of the companion merging galaxy \citep{Rodriguez2015}. 

This correlation between mass in the ICL and the assembly history of the system is consistent with being an extension into the larger mass regime of the trend found for stellar halos in MW-like galaxies \citep[e.g.,][]{Elias2018} and is also in agreement with findings in other simulations of MW-like and groups and cluster systems using the Horizon-AGN simulations \cite{Canas2020}. Similarly, several observational studies at low and intermediate redshift find a consistent pattern where higher ICL fractions are common in systems undergoing active mergers while lower fractions are characteristic in more passive or relaxed objects \citep{Jimenez-Teja2018, Jimenez-Teja2019, Jimenez-Teja2021, deOliveira2022, Dupke2022}. 

On the right panel of Fig.~\ref{fig2:f_ICL} we explore how TNG50 predictions compare to available theoretical and observational constraints to validate our systems. On the theoretical side, our results seem to agree, within the dispersion at a given mass, with semi-analytical models from \citeauthor{Contini2013}~(\citeyear{Contini2013}; shaded orange) and N-body + stellar tagging simulations of \citeauthor{Rudick2011}~(\citeyear{Rudick2011}; shaded cyan). 

Our results seem to also agree well with several observational constraints available in the literature (symbols with error bars), in particular, bearing in mind that the definition of ICL, band-width and depth of the observational data varies from system to system. We discuss in more detail the compilation of individual measurements in Appendix~\ref{app:obs} and also summarize the information in table ~\ref{tab:obs_data}. 

Most observational constraints seem to suggest that the ICL has less than $\sim 30\%$ of the light of the BCG, in good agreement with our results. Only NGC 5018 (red diamond), HCG 79 (highest purple triangle) and the Fornax cluster (pink triangle) have $f_{\rm ICL} > 0.3$. While in general we do not reproduce such large ICL fractions in our simulated systems, two of our groups have $f_{\rm ICL} \geq 0.3$. A careful look into the formation history of these systems (with FoF group IDs 4 and 12 in TNG50) revealed that their large ICL fractions can be attributed to two different factors. Group 4 presents a central galaxy that is under-massive compared to what is expected of its virial halo mass, such that when computing the ICL fraction it appears larger than other systems. Closer inspection revealed that this group had a rather late major merger (time $\sim 12.3$ Gyr) that significantly increased its total mass but has not yet propagated into the central galaxy yet. Reassuringly, this group has several large mass satellites that presumably will merge soon with the central bringing its mass into agreement with its halo mass, at which point we expect $f_{\rm ICL}$ to align with the lower values found in the sample. The second group, FoF group 12, is currently undergoing a major merger with a large satellite companion and the stripped material of this interaction is actively increasing the ICL mass without contributing yet to the central galaxy. Both of these examples are expected to decrease their $f_{\rm ICL}$ after the current interactions are settled. 

Our analysis of the groups in TNG50 also revealed the presence of a substantial in-situ stellar component within the ICL, in agreement with previous studies of stellar halos in the predecessor Illustris simulation \citep{DSouza2018, Elias2018} for smaller mass systems and a more recent study of cluster mass halos using TNG300 simulation by \citealt{Montenegro2023} (see Sec.~\ref{ssec:insitu} for a description of the in-situ definition). We highlight in Fig.~\ref{fig2:f_ICL} where the ICL fraction would fall if one only considers the accreted component (gray filled circles), indicating that the inclusion or not of this component does not strongly modify our results. A more detailed study of the in-situ ICL is deferred to a forthcoming paper.

%%%%%%%%%%%%%%%%%%%%%%%%%%%%%%%%%
\begin{figure}
	\includegraphics[width=\columnwidth]{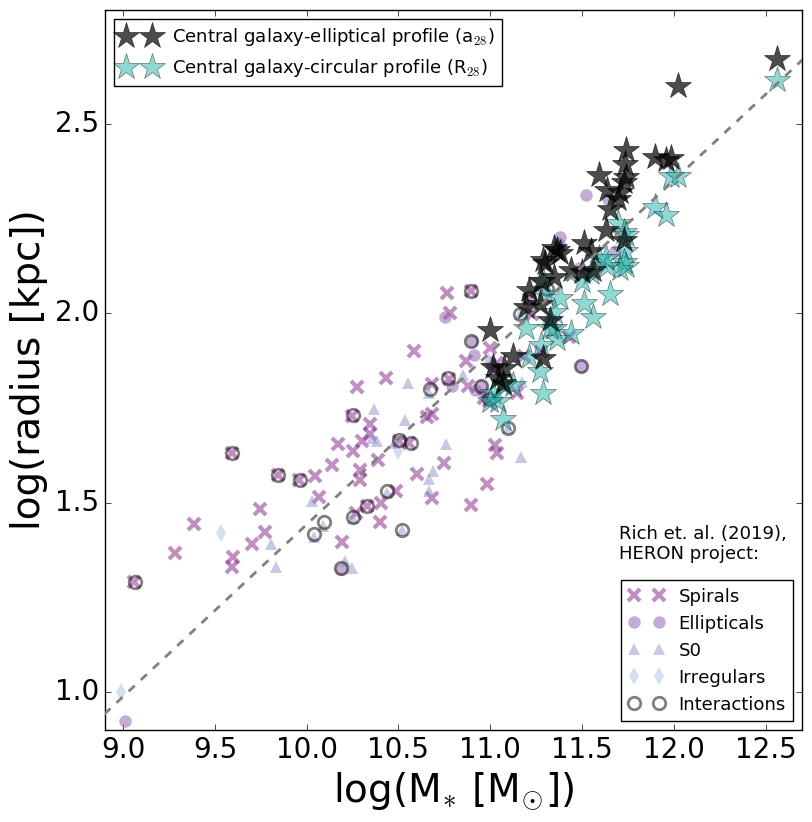}
    \includegraphics[width=\columnwidth]{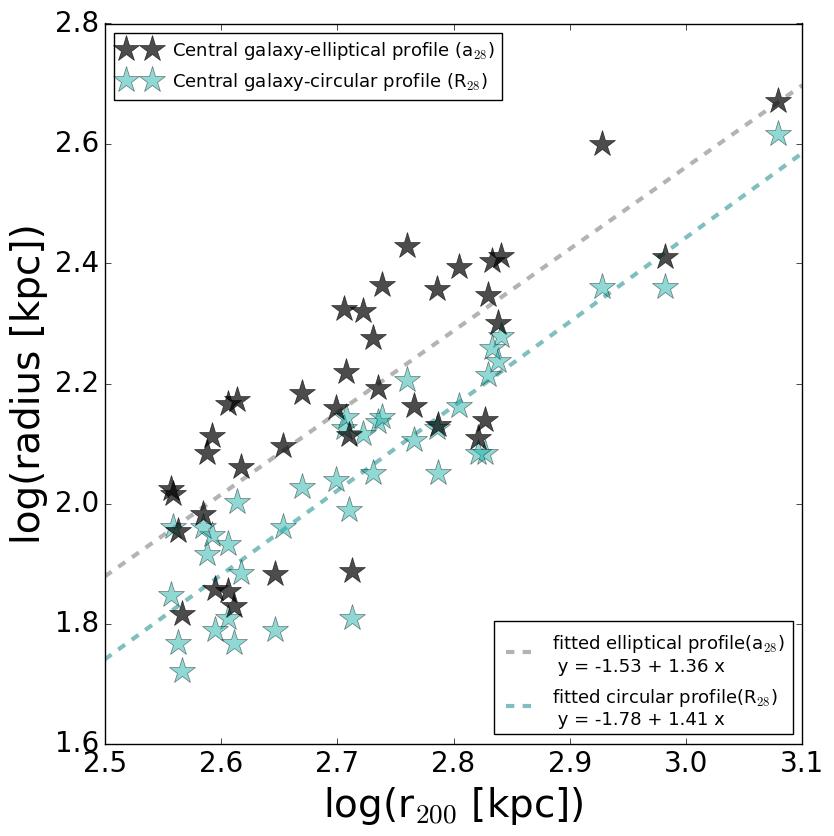}
    \caption{Size of the extended stellar halo measured by fitting ellipses to the surface brightness maps (black stars), and by measuring radius from radial binning of surface brightness maps (turquoise stars). Upper panel: Size-stellar mass relation, compared to observations from \protect\cite{Rich2019} (colored markers) and relation from \protect\cite{Munoz-Mateos2015} (gray dashed line). Lower panel: Size as a function of virial radius of the group. Grey and turquoise dashed lines correspond to measurements of semi-major axis of the best fit ellipse and radius of the circular profile, respectively.}
    \label{fig4:size_mass}
\end{figure}
%%%%%%%%%%%%%%%%%%%%%%%%%%%%%%%%%

\subsection{ICL radial extent}

The physical size at which the density profile of the ICL reaches a given surface brightness limit is an observable that may help constrain theoretical models on the formation of galaxies and their associated diffuse stellar component. In order to establish a fair comparison between observations and simulations, we randomly project our simulated groups and clusters generating their 2D surface brightness maps. Following \citet{Rich2019}, we measure the radius at which the surface brightness reaches $28$ mag/arcsec$^2$ in the r-band, using the available stellar particle magnitudes from TNG50.

We characterize the radial extension of the ICL in our simulated systems by defining two different radii, $R_{28}$ and $a_{28}$, an ellipsoid-base and a circle-based definition following common practice in the field. First, we measure the average surface brightness in circular bins and find the (circular) radius $R_{28}$ where the surface brightness is closest to 28 mag/arcsec$^2$. Second, we measure the surface brightness map with a 3 kpc $\times$ 3 kpc resolution and fit an ellipse to the regions of this map exhibiting a surface brightness of 28 magnitudes per square arcsecond (excluding any low surface brightness features that may result from interactions of subhalos). The semi-major axis of this ellipse was then used as a proxy of radial extension ($a_{28}$). Note that, on average, this surface brightness limit is predicted to occur at $\sim [0.2 \rm - 0.3] r_{200}$ in our systems, extending significantly beyond the central galaxy. Next generation surveys using Euclid or JWST are expected to reach several magnitudes fainter, mapping further out regions at $\sim [0.3 \rm - 0.9]r_{200}$ (see Fig.~\ref{fig:SB_circular} in Appendix ~\ref{app:sb}).

Fig.~\ref{fig4:size_mass} shows the size of our simulated systems as a function of the stellar mass of the central (top panel) and the virial radius of the group (bottom panel). As expected, the elliptical semi-major axis tends to be slightly larger than the circularly-averaged radius $R_{28}$, but they both trace similar trends. The ICL size follows a power law relation with the mass of the central galaxy, consistent with observational constraints available at the low mass end of our sample.

The top panel of Fig.~\ref{fig4:size_mass} shows this good agreement by including observational results for the stellar halo of galaxies in the HERON survey \citep{Rich2019} (magenta, purple and gray symbols) and also the fit provided in \cite{Munoz-Mateos2015} (gray dashed line). Our sample, while in agreement in the low mass end, seems to predict a slightly steeper increase in size with stellar mass for the highest mass objects than expected from the gray dashed line.

The bottom panel of Fig.~\ref{fig4:size_mass} indicates that the size of the ICL also shows a good correlation with the virial radius of the system, although with significant scatter. The gray and cyan dashed lines are best-fit linear relations (y = a + b x) between the logarithmic sizes of the ICL and the halos, characterized by the parameters a and b. For the elliptical radius $a_{28}$, we find the parameter values to be $a = -1.57$ and $b = 1.41$ with r.m.s scatter of $\sim 0.1$ dex, while for the circular radius $R_{28}$, the values are $a = -1.81$ and $b = 1.45$ with a scatter measured to be $\sim 0.1$ dex. The existence of a correlation between the radial extension of the ICL and the virial radius of the host halo demonstrates that the assembly of these two components is intertwined and reaffirms the validity of the ICL as observational tracer of the distribution of dark matter in these objects.

%%%%%%%%%%%%%%%%%%%%%%%%%%%%%%%%%%%%%%%
\begin{figure}
	\includegraphics[width=\columnwidth]{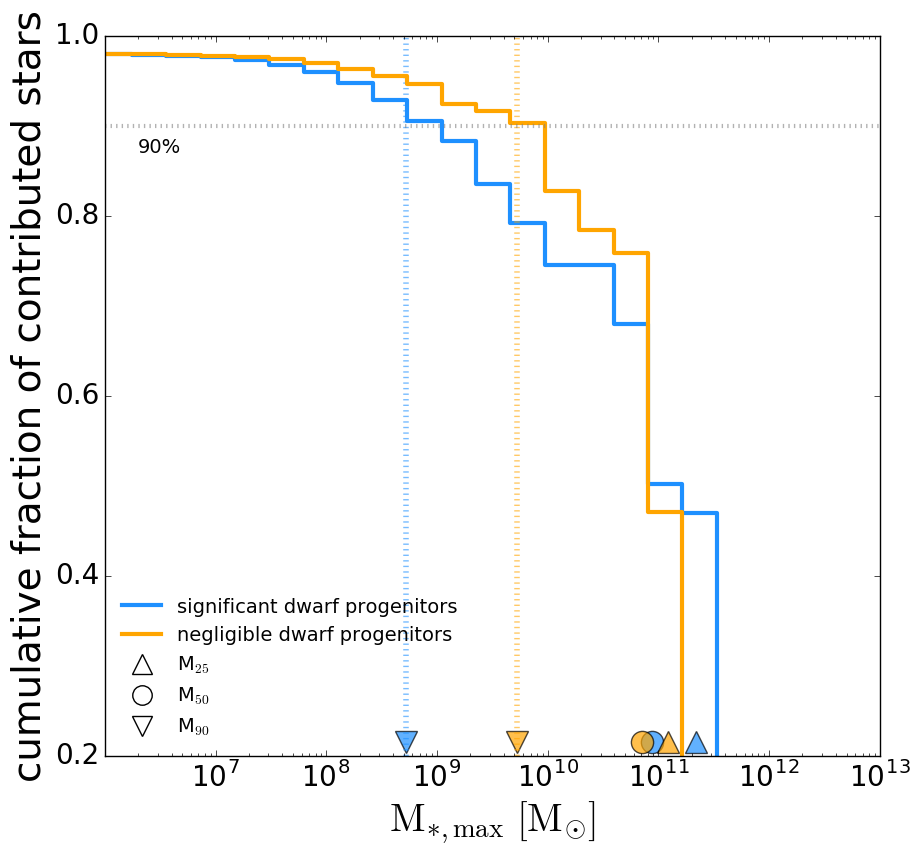}
    \caption{Example of two groups with different ICL progenitor mass functions. The vertical axis shows the cumulative fraction of stellar mass deposited in the ICL by galaxies with maximum stellar mass $>\rm M_{*,\rm max}$. Markers on the x-axis correspond to the measured M$_{25}$ (triangle), M$_{50}$ (circle), and M$_{90}$ (pointing down triangle), and horizontal and vertical lines correspond to the measurement of the M$_{90}$ for the groups highlighted here. The group in orange has an ICL contributed mostly by massive progenitors with $M_{*, \rm max}\sim 10^{10} \; \rm M_\odot$ while the one in blue has significant contributions from lower mass dwarf galaxies with $M_{*, \rm max} \sim 5 \times 10^{8} \; \rm M_\odot$, as shown by their different $M_{90}$ values.}
    \label{fig6:prog_example}
\end{figure}
%%%%%%%%%%%%%%%%%%%%%%%%%%%%%%%%%%%%%%%

%%%%%%%%%%%%%%%%%%%%%%%%%%%%%%%%%%%%%%%
\begin{figure*}
    \includegraphics[width=2\columnwidth]{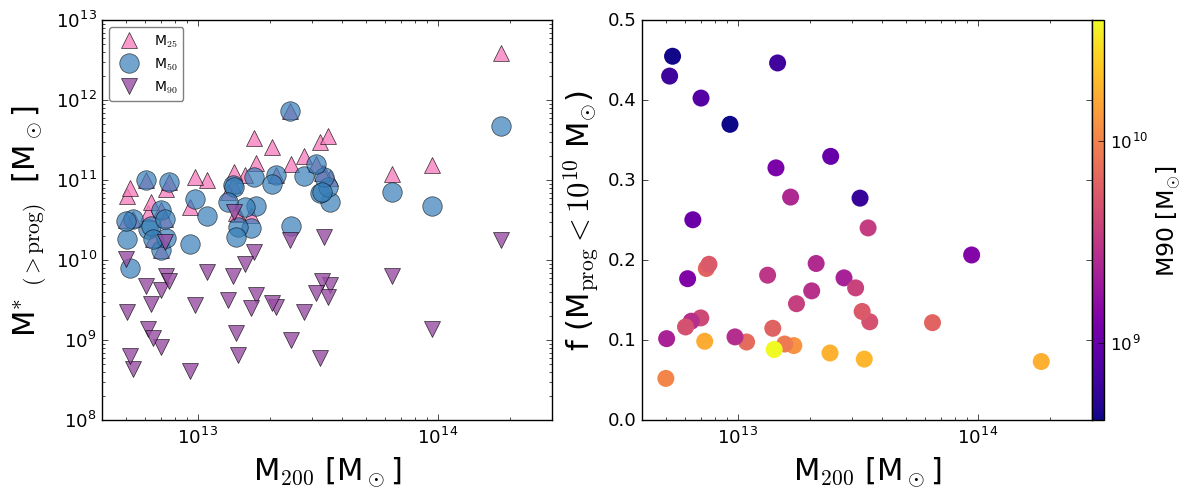}
    \caption{{\it Left panel:} Stellar mass of the progenitors that brought a percentile of mass to the ICL as a function of virial mass of the groups. Pink triangles, blue circles, and purple pointing down triangles correspond to M$_{25}$, M$_{50}$, and M$_{90}$, respectively. Note the large scatter in $M_{90}$ indicating a varied contribution of dwarf galaxies to the build up of the ICL.  {\it Right panel:} fraction of the accreted ICL that was contributed by (dwarf) galaxies with stellar masses below $M_{*,\rm max} = 10^{10}$ \msun. Symbols are colored by M$_{90}$ for each group (from the left panel) and highlight the good correspondence between the two metrics. At similar halo mass, some groups have about half of their ICL built by low mass galaxies with $M_{*,\rm max} = 10^{10}$ \msun\; while others show less than 10\% contribution from these dwarfs.} 
    \label{fig7:progenitors}
\end{figure*}
%%%%%%%%%%%%%%%%%%%%%%%%%%%%%%%%%%%%%%%

%%%%%%%%%%%%%%%%%%%%%%%%%%%%%%%%%%%%%%%
\begin{figure*}
	\includegraphics[width=2.05\columnwidth]{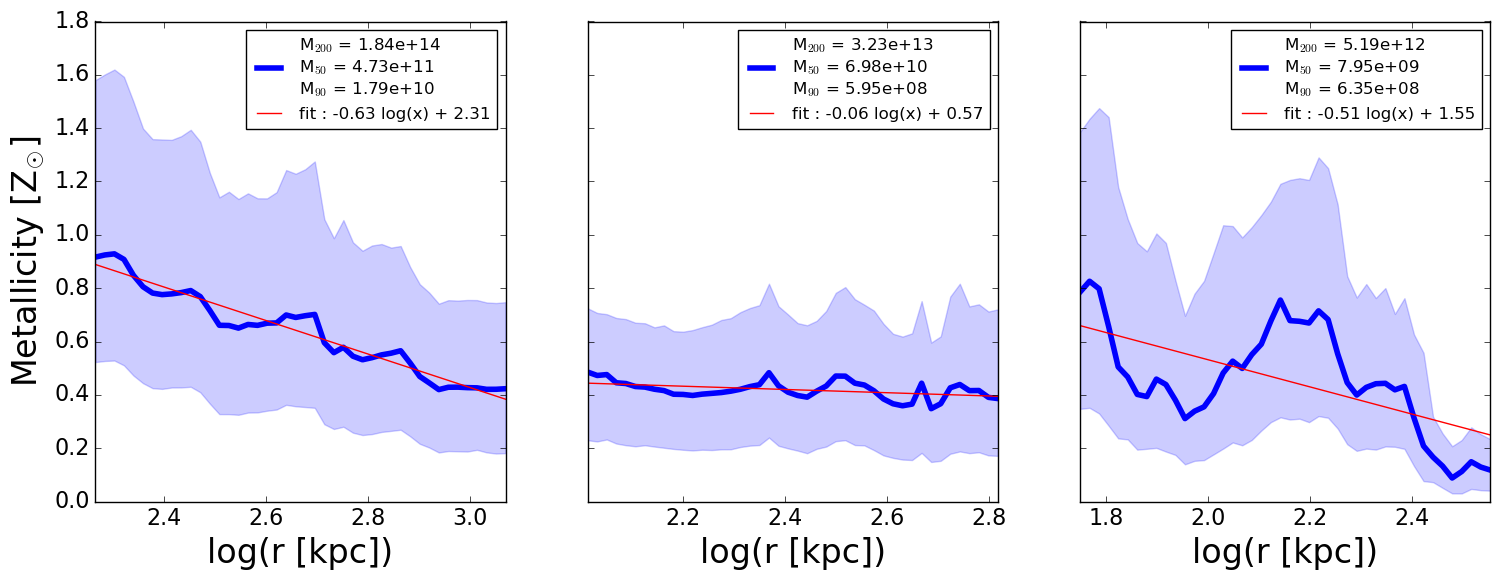}
    \caption{Illustration of the metallicity profiles of three randomly selected groups from our sample, with the median and 25-75 percentile depicted by the blue line and shaded region, respectively. The best-fit linear relation (in log) is highlighted in red and quoted in each panel. Most systems show a negative metallicity profile with radius, but the slope changes substancially from system to system.}
    \label{fig8:metal_profile}
\end{figure*}
%%%%%%%%%%%%%%%%%%%%%%%%%%%%%%%%%%%%%%%

%%%%%%%%%%%%%%%%%%%%%%%%%%%%%%%%%%%%%%%
\begin{figure*}
    \includegraphics[width=2.05\columnwidth]{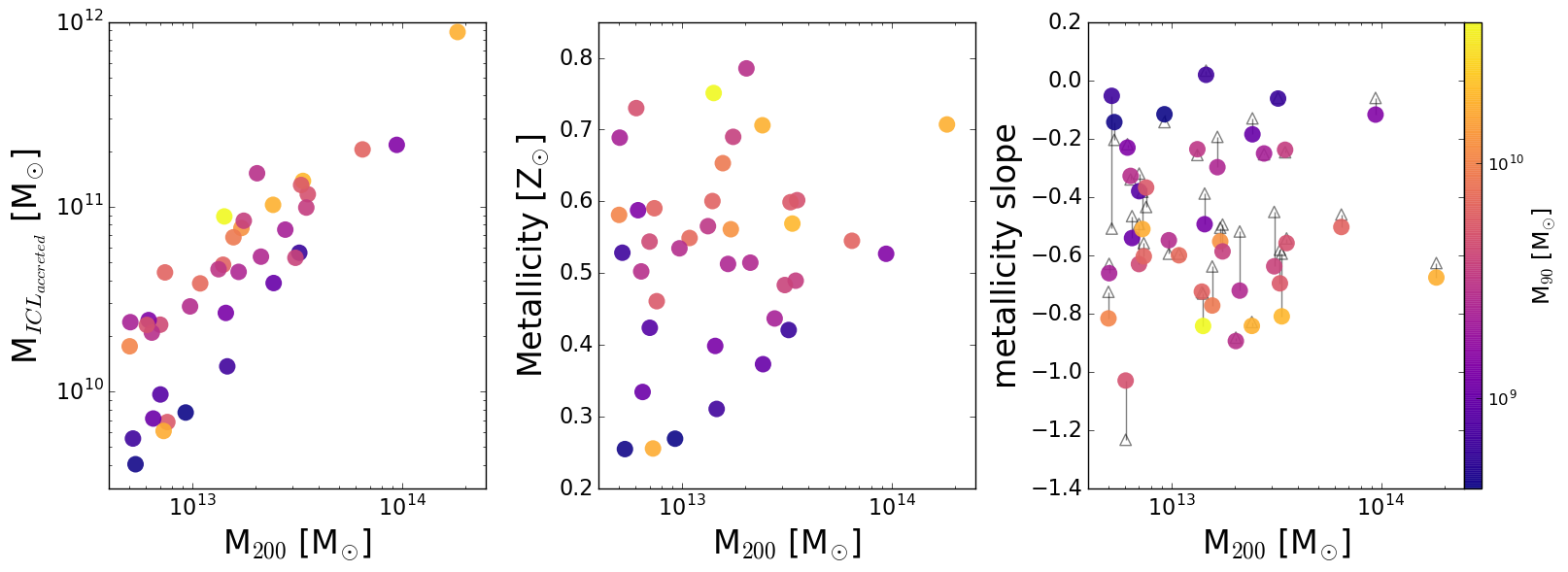}
    \caption{Left panel: stellar mass of the accreted component in the ICL as a function of the virial mass. Middle panel: median metallicity of the accreted stars in the ICL as a function of the virial mass. Right panel: the slope of the metallicity profile as a function of virial mass for the merged component, with black triangles representing the results for all the stellar components. All panels are color-coded by the $M_{90}$ of each group. Systems where dwarf galaxies play a role in building the ICL (low $M_{90}$) tend to have lower ICL mass, lower overall metallicity and shallower metallicity profiles than systems with accretion dominated by more massive galaxies.}
    \label{fig9:metal_slope}
\end{figure*}
%%%%%%%%%%%%%%%%%%%%%%%%%%%%%%%%%%%%%%%

\section{The progenitors of the ICL in groups and clusters}\label{Prog}

A myriad of disrupted satellites are expected to contribute to the build up of the ICL as a consequence of the hierarchical formation scenario, making it a unique probe of the particular assembly history of a given object. Yet, while the individual formation histories may vary from halo to halo, common trends arise that may be used to infer details of the formation and assembly of a given object from the properties of their diffuse stellar component. For instance, in the case of MW-mass galaxies, simulations suggest that the properties of the stellar halos are dominated by one or two most massive progenitors, which naturally explains the correlation between stellar halo mass and metallicity \citep[e.g.,][]{DSouza2018, Deason2016}. We explore in what follows what are the typical building blocks of the ICL in more massive objects like groups and clusters, and what observational signatures may be useful to decode details of their assembly. 

Fig~\ref{fig6:prog_example} shows the fractional contribution of stars to the ICL of two different groups (orange and blue curves) that come from progenitors of a given stellar mass $M_{*, \rm max}$. Both groups are selected to have comparable mass, $M_{200} \sim 3.2 \times 10^{13}$ $\rm M_\odot$, but different accretion histories. Here we choose to characterize the mass of the progenitors using their maximum stellar mass $M_{*, \rm max}$, and we build the cumulative histogram of stellar particles brought in by progenitors more massive than a given $M_{*, \rm max}$. The group depicted in orange corresponds to a case where most of the mass in the ICL is contributed by relatively massive galaxies while the group shown in blue allows for significant contribution of progenitors in the regime of dwarf galaxies. 

We quantify this by means of $M_{25}$, $M_{50}$ and $M_{90}$, which are defined as the mass of the progenitors in such cumulative distribution contributing $25\%$, $50\%$ and $90\%$ of the mass in the accreted ICL. For illustration, we highlight $M_{90}$ in Fig.~\ref{fig6:prog_example} with vertical dotted lines and an inverted triangle, computed as the intersection of the cumulative distribution of accreted stars in each group and the $0.9$ horizontal line. $M_{50}$ and $M_{25}$ are also denoted by a circle and a triangle for each group along the horizontal axis. The larger contribution of dwarf galaxies to the build up of the ICL in the group shown in cyan is now clearly shown by its lower mass $M_{90} = 5.2 \times 10^{8}$ M$_{\odot}$ compared to $M_{90} = 5.2 \times 10^{9}$ M$_{\odot}$ in the case of the group in orange. Differences for $M_{50}$ and $M_{25}$ between these two groups are smaller, but systematic.

The typical progenitors of the ICL for the full sample of groups and clusters are shown in the left panel of Fig.~\ref{fig7:progenitors}. Our results suggest that the predominant contributors to their ICL are massive galaxies with $M_{50} \geq 10^{10}$\msun, regardless of the host halo mass in this somewhat narrow $M_{200}$ range. This is in agreement with previous works in the literature which identify $\rm M_{*} \sim 10^{10}$-$10^{11}$\msun\; as the main progenitors of the ICL in groups \citep[e.g.,][]{Contini2013, Contini2019, Montes2018, Montes2021}. However, a comparison of the $M_{90}$ values uncovers a larger variability of contributions of dwarf galaxies from halo to halo, with some systems having $M_{90} \sim 10^{10}$\msun\; --and therefore having negligible mass brought in by dwarf galaxies with $M_* \sim 10^9$\msun\; and below-- while other systems show a more significant contribution by dwarfs with typical $M_{90} < 10^9$\msun. 

This is further illustrated in the right panel of Fig.~\ref{fig7:progenitors}, which depicts the fraction of ICL mass contributed by galaxies with $M_{*,\rm max} < 10^{10}$ \msun\; as a function of the halo virial mass. As anticipated, these fraction can be quite different from system to system, even in the case of similar halo masses. For instance, for groups with $M_{200} \sim 10^{13}$\msun, the fraction of mass brought into the ICL by low mass galaxies with $M_{*,\rm max} < 10^{10}$ \msun\; varies from $10$\% to $40$\%, these extremes being examples of systems where dwarf galaxies play little to a significant role, respectively. Symbols in the right panel of Fig.~\ref{fig7:progenitors} are colored according to the previously introduced $M_{90}$ and highlights the good correlation between both indicators: systems with a substantial contribution from low mass galaxies have a high fraction of stars brought in by progenitors with $M_{*,\rm max} < 10^{10}$ \msun\; and a low $M_{90}$ value while systems built up mostly by large galaxies have a small fraction $f$ and large $M_{90}$ values.

This opens up the possibility to use different observables to attempt determine the kind of accretion history of a halo given the observed properties of their ICL. One such key observable is the metallicity profile. Fig~\ref{fig8:metal_profile} illustrates the metallicity profile for three representative groups with differing formation histories and virial masses (as indicated by the legends). The blue curve in each panel depicts the median metallicity of the stellar particles in the ICL at each radius, while the blue shaded region represents the 25th and 75th percentiles. A linear fit of the form $Z = a\rm log(r)\; +\; b$ is in general a roughly good description of our profiles (see red lines) and allow us to quantify for each object the slope of the metallicity profile ($a$) and its intercept ($b$).

Fig.~\ref{fig9:metal_slope} shows interesting correlations in the ICL observables that imprinted by the kind of progenitors that built each individual object. At a given host halo mass, systems with a larger contribution from dwarfs (blue-ish points) show less total mass in the ICL (left panel), a lower average metallicity (middle panel) and also flatter metallicity profile slopes (right-hand panel). Systems with mostly massive progenitors (red colors) tend to show strongly declining metallicity profiles. Not shown here for brevity, the metallicity profile intercepts correlates well with the median metallicity. In all panels points have been color-coded by $M_{90}$ and only the accreted component of the ICL is being considered here. However, our main results would not substantially change if, instead, we would show the total ICL (in-situ plus accreted, gray triangles on the right panel). We have explicitly checked that similar trends exist with the stellar age profiles, with most systems displaying decreasing age profiles with radius, and some showing considerably flatter age distributions. However ages seemed less well correlated with the assembly history of the group and cluster than the information provided by metallicity.

%%%%%%%%%%%%%%%%%%%%%%%%%%%%%%%%%%%%%%%%%
\begin{figure*}
	\includegraphics[width=2.05\columnwidth]{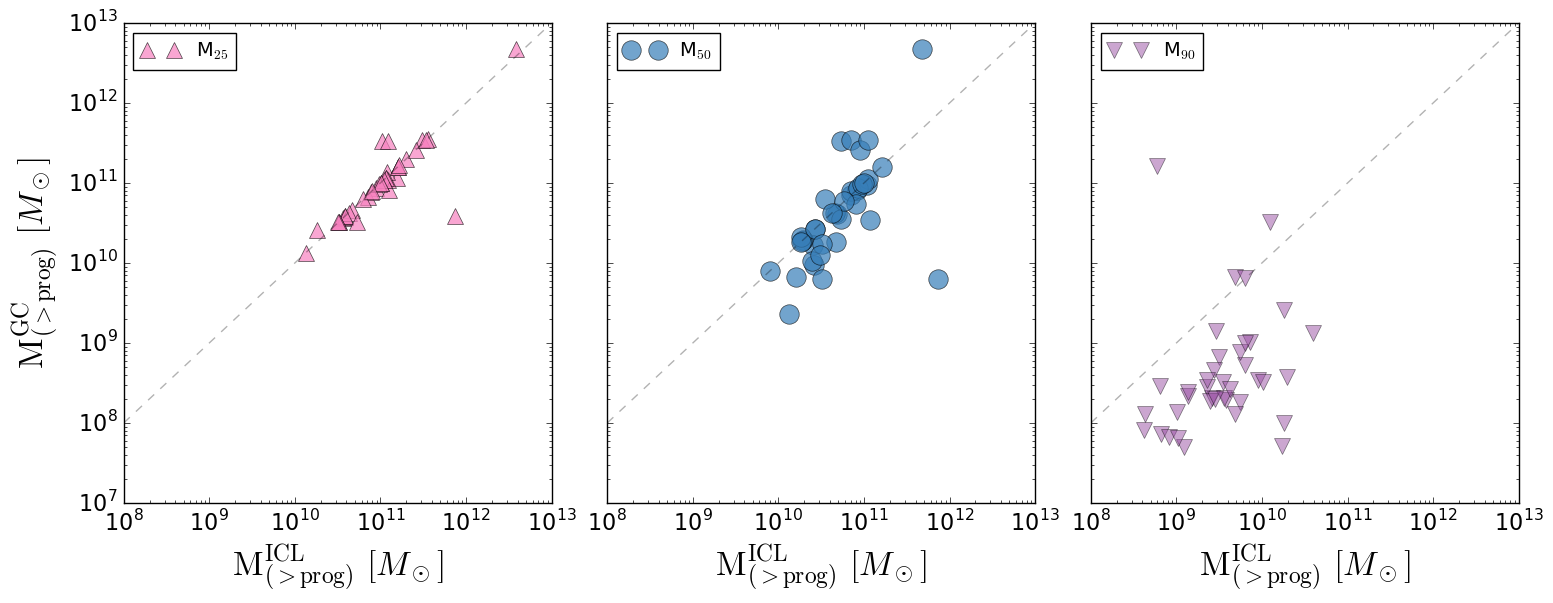}
    \caption{Typical mass of progenitors that build the ICL (x-axis) and the intracluster-GCs (y-axis) as quantified by M$_{25}$ (left), M$_{50}$ (middle), and M$_{90}$ (right) of each population. The gray dashed line corresponds to the one to one relation. Our findings indicate that GCs can serve as a good tracers of the most massive contributors to the ICL, although it is important to note that the contribution from dwarf galaxies may be over-represented when looking at GCs in comparison to the ICL (right panel).}
    \label{fig10:GC_vs_ICL}
\end{figure*}
%%%%%%%%%%%%%%%%%%%%%%%%%%%%%%%%%%%%%%%%%

The ICL mass and metallicity should then be considered a valuable tool to reconstruct details on the past merger history of systems assembled within $\Lambda$CDM. Similar correlations on amount of mass, overall metallicity and metallicity gradients with the diffuse component build-up have been found in the regime of stellar halos for MW-like galaxies \citep{DSouza2018,Monachesi2019}. Our results extend upwards the range of halo masses over which these correlation are expected, including now the regime of groups and clusters.  

%%%%%%%%%%%%% GCs section %%%%%%%%%%%%
\section{Tracing the formation history of the ICL through globular clusters} 
\label{GCs}

We investigate how well the GCs can trace the ICL in our systems using the GC catalog presented in \citet{GCCat}. As detailed in section~\ref{method4}, we use only intra-cluster GCs (ICGCs)---GCs not gravitationally bound to any structure according to {\sc SUBFIND} and that satisfy the cleaning criteria in section~\ref{method2}. As briefly discussed in \citet{GCCat}, the GC tagging model predicts that the ICGC component forms via the tidal stripping of GCs from their host galaxies as they interact with their new host environment after infall. The buildup of ICGCs is thus, in a similar fashion to the ICL, a result of the hierarchical assembly of their host systems. The GC catalog follows the population of surviving GCs with mass $M_{\rm GC} > 7 \times 10^{3}$ \msun, and while we do not make the distinction between ``red" and ``blue" GCs as in \citet{GCCat}, their Fig.5 shows that the mass density of the simulated ICGC is in good agreement with available observational constraints.

As in the case of the ICL, we find a large spread on the distribution of progenitors that build the the intra-cluster GC component in our groups. Using a similar concept to that introduced for $M_{25}$, $M_{50}$ and $M_{90}$, we define similar quantities but considering only the stellar mass that brought in the $25\%$, $50\%$ and $90\%$ of the mass in GCs in the ICGC component. Fig.~\ref{fig10:GC_vs_ICL} compares the progenitors for the ICL and the GCs. We find that GCs can be very good tracers of the most massive contributors to the ICL, in particular $M_{25}$ and $M_{50}$ for the stars or GCs in the diffuse component are very similar, indicating that the progenitors that build up to half of the ICL also bring along about half of the GCs in the ICGC. However, we find a clear bias in the contributors as quantified by $M_{90}$: taking the progenitors that bring 90\% of the GCs typically leads to smaller masses than the progenitors contributing 90\% of the stellar mass in the ICL (left panel Fig.~~\ref{fig10:GC_vs_ICL}). 

This means that when using GCs to reconstruct the merger history of groups and clusters, one should keep in mind that the contribution from dwarf galaxies will be over-represented compared to the contribution of the same dwarfs to build the diffuse light component. This result can be intuitively understood from the different scaling of GCs and stellar mass with the halo mass. In the considered GC model, the mass of GCs scales as a power-law of halo mass \citep{GCCat} while abundance matching results suggest a more steep decrease in the efficiency of low mass halos to form stars compared to more massive galaxies like the MW (abundance matching is better described by a double power-law). As a result, dwarfs contribute {\it fractionally} more GCs than stars to the diffuse components. Interestingly, as found for the case of the ICL, there is a sizable range in the typical contributor of the ICGC. For some of our GC systems, $M_{90}$ is found to be galaxies with $M_* \sim 10^{10}$\msun, indicating little role played by low mass galaxies. However, there are also cases with $M_{90} \sim 5 \times 10^7$\msun, indicating dwarfs with masses typical of dSph galaxies playing a role.  These differences in the accretion histories of these objects should remain imprinted in the chemical properties of their intra-cluster GC component and offer an avenue to constrain formation histories of massive hosts in cases where measurements of diffuse light becomes too challenging or implausible.

\section{Conclusions}\label{sec:conclusion}

Using the cosmological hydrodynamical simulation TNG50, we have conducted a comprehensive study of two luminous tracers of dark matter in the outskirts of halos: the intra-cluster light (ICL) and the globular cluster (GC) population. We select all groups and clusters with $M_{200}>5 \times 10^{12}$\msun\; resulting in a sample of $39$ host halos that span the virial mass range $M_{200} = [0.5 \rm - 20] \times 10^{13}$\msun\; and sample a wide range of formation histories. We use the catalog of GCs tagged to the TNG50 groups and clusters presented in \citet{GCCat}, allowing one of the first explorations of both luminous tracers in high density environments. We focus on the study of possible imprints of these individual accretion histories on observable properties of the ICL and GCs populations. Our findings can be summarized as follows:

\begin{itemize}
    \item TNG50 predictions for the fraction of light in the ICL, $f_{\rm ICL}$, are in reasonable agreement with current observational constraints for groups and low mass clusters. In addition, $f_{\rm ICL}$ is predicted to increase with virial halo mass, albeit with significant dispersion. We find that this dispersion is well correlated with the assembly history in our systems causing variations in the dynamical state of the system and the ICL. Objects with more recent major mergers tend to show an excess of ICL compared to older or earlier assembled systems of comparable mass.  
    
    \item There is a good correlation between the radial extent of the ICL (as measured by the radius where the surface brightness profile reaches $28$ mag/arcsec$^2$) and the stellar mass of the central galaxy. This size--mass scaling relation for the ICL seems to agree with the one found observationally for lower mass galaxies as part of the stellar halos in the HERON survey \citep{Rich2019}. The radial extent of the ICL also shows a power-law relation to the virial radius of the halo, offering an observational means with which to constrain $r_{200}$ independent of abundance matching methods. The typical r.m.s scatter in the relation between ICL-size and virial radius is $\sim 0.1$ dex. 

    \item The mass of the ICL is deposited by a wide range of progenitors. Some of our systems show sizable contributions from dwarf galaxies with $M* < 10^9$\msun, while others are built up almost entirely by systems as or more massive than the Milky Way. We find that the metallicity and metallicity profiles of the ICL retain information on these different assembly histories: halos with a higher contribution from dwarf galaxies generally have shallower metallicity profiles, overall lower average metallicities and also a lower amount of mass/light in the ICL. These findings are different from the common interpretation of flat metallicity profiles as evidence of large major mergers \citep{Krick2007, DeMaio2015, DeMaio2018}. In our systems, it is 
    the incidence of low mass galaxies the main responsible for flattening up the $z$-profiles. A larger role of dwarf galaxies in the build up of the ICL is also manifested by an overall lower metallicity content compared to systems built up by more massive satellite contributors. 

    \item Finally, our study highlights that intra-cluster globular clusters (ICGCs) may also serve as a valuable tool to reconstruct the assembly history of the halos. We find that lower mass dwarfs tend to make more significant contributions to the ICGC population than to the stars in the ICL, a bias that must be taken into account when reconstructing assembly histories based on metallicity of GCs or ICL. This can be understood as a consequence of the single power-law relation of GCs with halo mass, which gives higher weight to the low mass halos compared to the double-power law expected for the stars in abundance matching relations.
    
\end{itemize}

Several of the trends studied here represent an extension of properties highlighted for the stellar halos of galaxies like the Milky Way, including trends on amount \citep{Pillepich2014, Elias2018}, metallicity \citep{DSouza2018,Monachesi2019} and radial extension \citep{Rich2019} and their link to the assembly history of each halo. Encouragingly, the ICL in groups and clusters are more massive/brighter than in systems like the Milky Way, offering an advantage from the observational point of view to study this diffuse component. The increased sensitivity of upcoming observations with JWST and Euclid VIS Deep survey promise to reach $\sim 31$ mag/arcsec$^2$ levels in a few hours integration time. On median, for our simulated systems, this would result in detections for the ICL component all the way out to $\sim 0.7 r_{200}$, opening a new window to explore the predictions of hierarchical assembly expected for galaxies within $\Lambda$CDM. 

\section*{Acknowledgements}

We gratefully acknowledge the contributions of several individuals who supported this research. We would like to express our appreciation to Mireia Montes for generously providing the valuable data from \citep{MontesReview2022}. We are also thankful to Nima Chartab for engaging in insightful conversations and offering valuable inputs. NA, LVS, and JED would like to acknowledge the financial support received from the NASA ATP-80NSSC20K0566,  NSF-CAREER-1945310 and NSF-AST-2107993 grants.

%%%%%%%%%%%%%%%%%%%%%%%%%%%%%%%%%%%%%%%%%%%%%%%%%%
\section*{Data Availability}

This paper is based on halo catalogs and merger trees from the Illustris-TNG Project \citep{Nelson2019}. These data are publicly available at \href{https://www.tng-project.org/}{https://www.tng-project.org/}. The realistic GC catalogs are available to the public. They can be downloaded from: \href{www.tng-project.org/doppel22}{www.tng-project.org/doppel22}.



\bibliographystyle{mnras}
\bibliography{example} % if your bibtex file is called example.bib

%%%%%%%%%%%%%%%%%%%%%%%%%%%%%%%%%%%%%%%%%%%%%%%%%%

%%%%%%%%%%%%%%%%% APPENDICES %%%%%%%%%%%%%%%%%%%%%

\appendix

\section{Effect of ICL definition} \label{App A}

\begin{figure*}
	\includegraphics[width=2.05\columnwidth]{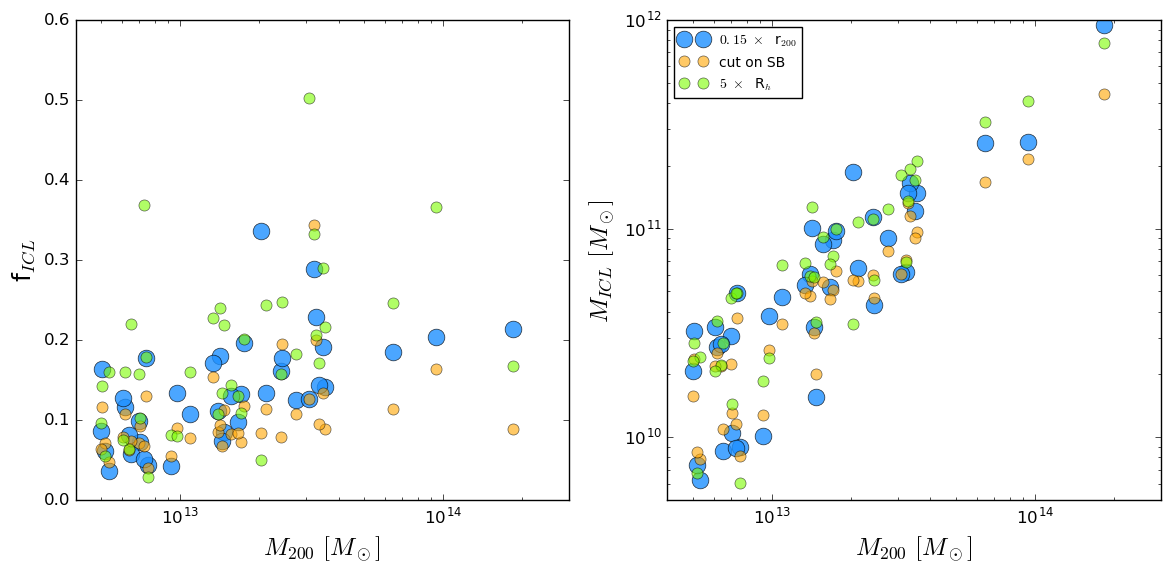}
    \caption{Left panel: ICL fraction as a function of virial mass. Right panel: stellar mass content in the ICL as a function of virial mass. The different colors correspond to various definitions of ICL.}
    \label{fig:app1}
\end{figure*}

As mentioned in section~\ref{amount} there are various definitions of ICL in the literature. Here we examine the effect on the predicted ICL mass of adopting a few of these different definitions. The methods used to define ICL are as follows; considering stellar particles that do not belong to any subhalos in the range $0.15 \times \rm r_{200} < \rm r < \rm r_{200}$ (shown by blue circles), compared to stars in $5 \times \rm r_{h_{*}} < \rm r < \rm r_{200}$ where $\rm r_{h_{*}}$ is the stellar half mass radius of the central galaxy (shown by green circles), and stars in range $\rm r_{25} < \rm r < \rm r_{200}$ where $\rm r_{25}$ is the radius where the surface brightness (SB) reaches 25 mag/arcsec$^2$ (results are shown by orange circles). The right panel of Fig.~\ref{fig:app1} shows the mass in the ICL, while the left panel shows ICL fraction as a function of virial mass of the group for our sample of 39 halos in TNG50. Our results suggest that, given the dispersion in the predicted ICL amount at a given virial mass, the changes in the definition will not significantly affect the results.

\section{The GC tagging technique} 
\label{app:gcs}

Below is a summary of the method introduced in \citet{GCCat} to tag GCs into the TNG50 simulations. We add GC particles to all selected galaxies in TNG50 that interact with the 39 most massive groups ($M_{200} > 5 \times 10^{12}\; \rm M_\odot$). The procedure is done at their time of infall, when we select the full set of DM particles that follow a given energy distribution. The energy distribution is calculated by assuming that the DM halos of the selected galaxies conform to an NFW profile \citep{Navarro1996}:
\begin{equation}
    \rho_{\rm NFW}(r) = \frac{\rho_{\rm NFW}^0}{(r/r_{\rm NFW})(1 + r/r_{\rm NFW})^2},
\end{equation}
that is fit to each galaxy at infall following \citet{Lokas2001}. We assume a scale radius, $\rm r_{NFW} = \rm r_{max}/ \alpha$ where $\rm r_{max}$ is the radius of maximum circular velocity calculated from the dark matter particle distribution in the simulation and $\alpha = 2.1623$ \citep{Bullock2001}.

Subsequently, GCs are assumed to follow a Hernquist profile \citep{Hernquist1990}:
\begin{equation}
    \rho_{\rm HQ}(r) = \frac{\rho_{\rm HQ}^0}{(r/r_{\rm HQ})(1 + r/r_{\rm HQ})^3}.
\end{equation}

\noindent
with scale radius $r_{\rm HQ}$ is dependent upon whether the GC is ``red'' or ``blue''. The ``red'' component is intended to be representative of a more radially concentrated metal rich GCs, while  a ``blue'' component refers to a more radially extended and metal poor population of GCs. Following \citep{Ramos-Almendares2020}, it is assumed that $r_{\rm HQ} = \beta r_{\rm NFW}$, where $\beta_\mathrm{red} = 0.5$ and $\beta_\mathrm{blue} = 3.0$. $\rho_{\rm HQ}$ is fit such that the number of GC candidates are maximized. 

Next, the distribution function is calculated for the NFW halo, the blue GCs, and the red GCs following \citet{Binney2008}:
\begin{equation}
    f_{i}(\epsilon) = \frac{1}{8\pi}\bigg[ \int_{0}^{\epsilon} \frac{\rm d^2\rho_i}{\rm d\psi^2} \frac{\rm d \psi}{\sqrt{\epsilon - \psi}} + \frac{1}{\sqrt{\epsilon}} \bigg(\frac{\rm d\rho_i}{\rm d\psi}\bigg) \bigg|_{\psi = 0}\bigg] ,
    \label{eqn:distfunc}
\end{equation}
where $\rho_i$ is the density profile of i = (DM, Red GCs, Blue GCs), $\Psi$ is the relative gravitational potential, and $\epsilon$ is the relative energy. In bins of relative energy, a fraction $\rm f_{HQ, i}/\rm f_{DM}$ of the DM particles are selected to be GCs, and the final set of GC candidates are defined to be within a cutoff radius of $\rm r_h/3$, where $\rm r_h$ is the total half-mass radius of the galaxy, following \citet{Yahagi2005}. 

The method assumes a power-law relation between the total GC mass and the halo mass of each galaxy whose normalization is calibrated such that it reproduces the observed relation at the present-day \citep{Harris2015}. Only halos with $M_{200}>10^{11}\; \rm M_\odot$ participate in the calibration process and properties of lower mass objects can be considered a prediction. More specifically, the resulting coefficients for the $M_{GC}$ - $\rm M_{halo}$ relation at infall are:

\begin{equation}
    M_{\mathrm{GC}, z = 0} = a M_{\rm halo, z = 0}^b ,
     \label{eq:calibration1}
\end{equation}

\noindent
where $\rm a = 2.6\times 10^{-8}$  and $4.9\times 10^{-5}$ for red and blue GCs respectively, with $\rm b = 1.2$ and $0.96$ for red and blue GCs. Similarly to \citet{Harris2015}, $\rm z = 0$ halo masses are calculated assuming abundance matching parameters from \citet{Hudson2015}. 

The calibration for GC mass is made using the amount of GC stripping by $\rm z = 0$: $\rm f_{bound} = \rm N_{candidates, z = 0}/\rm N_{candidates, infall}$. Where a GC candidate is considered (still) bound to a galaxy if its corresponding DM particles are identified as part of the galaxy via {\verb'SUBFIND'}. Assuming that at infall, the relation between $\rm M_{GC}$ and $\rm M_{halo}$ still follows a power law:
\begin{equation}
    M_{\rm GC, inf} = \frac{1}{f_{\rm bound}} M_{\mathrm{GC}, z = 0} = a_{\rm inf} M_{\rm halo, inf}^{b_{\rm inf}} ,
    \label{eq:calibration2}
\end{equation}
with the coefficient and exponent found to be $\rm a_{\mathrm{inf}}$ = $2.6 \times 10^{-7}$ and $7.3 \times 10^{-5}$ and $b_{\mathrm{inf}}$ = 1.14 and 0.98 for red and blue GCs respectively. 

Finally, the number of all identified GC candidates with energy consistent with the distribution of red and blue GCs is in most cases larger than the observed number of GCs around such galaxies. Subsequently, the GC candidates at infall for each galaxy are subsampled to obtain a realistic number of GCs. GC luminosities are selected---that are converted to masses assuming a mass to light ratio of 1 in the z-band---from Gaussian luminosity functions with a galaxy-luminosity-dependent dispersion (Gaussian luminosity functions with a dispersion dependent on galaxy luminosity are used to select GC luminosities, which are then converted to masses, assuming a mass-to-light ratio of 1 in the z-band). The resulting individual GC masses, $\rm m_{GC}$ are in the range $7\times 10^3 \leq \rm m_{GC} \leq 5\times 10^6$ M$_{\odot}$, consistent with observational constraints from \citet{Jordan2007}.

%%%%%%%%%%%%%%%%%%%%%%%%%%%%%%%%%%%%%%%%%%%%%%%%%%%%%%%%%%%%%%%%%%
\section{Surface brightness profiles}
\label{app:sb}

\begin{figure}
	\includegraphics[width=\columnwidth]{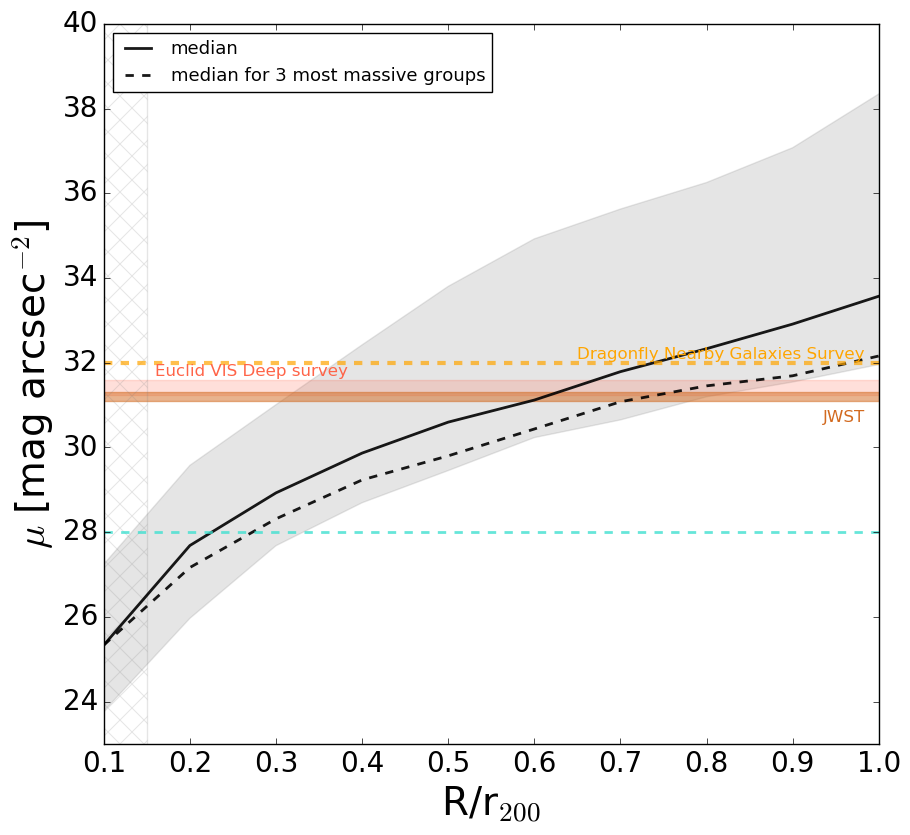}
    \caption{Surface brightness profiles of our sample of groups and clusters. The black line represents the median profile for the entire sample, while the dashed black line corresponds to the median profile of the three most massive groups. The shaded region illustrates the dispersion across all groups in our sample. The gray hatched region indicates the extent of the BCG.}
    \label{fig:SB_circular}
\end{figure}

In order to establish a fair comparison between observations and simulations, we randomly project our simulated groups and clusters generating their 2D surface brightness maps. The results are presented in Fig.~\ref{fig:SB_circular}, where the solid black line represents the median surface brightness for our entire sample in circular radii bins. Because we span a range of host halo masses, we normalize the horizontal axis to the virial radius of each system. The shaded grey region illustrates the whole range of surface brightness covered by our entire sample. This variation in profile is given by two factors: object-to-object differences in assembly history and mass in the ICL and, also important, the range of host halo masses being included. For illustration, the median surface brightness including only the three most massive clusters with $M_{200} \sim 10^{14}$\msun\; is shown by the dashed black line, which is systematically $\sim 1$ magnitude brighter that considering the whole sample. The area being shaded by hatching for $R/r_{200} < 0.15$ separates the inner regions attributed to the BCG and not considered part of the ICL.  

The turquoise horizontal dashed line indicates the 28 mag/arcsec$^2$ limit at which we measure the radius of each group. For comparison, we have included estimates of the surface brightness limits for the James Webb Space Telescope (based on the study of ICL in SMACS 0723 by \citealt{Montes2022}\footnote{\label{footnote_obslim}These limits correspond to a sky fluctuation of 3$\sigma$ in an area of 10 × 10 arcsec$^2$.}), the Euclide VIS Deep survey\citep{Borlaff2022}$^{\ref{footnote_obslim}}$, and the Dragonfly Telephoto Array's deep nearby galaxy survey (which provides g-band surface brightness profiles down to 31-32 mag/arcsec$^2$, see \cite{Merritt2016} for applications to stellar halo studies). We predict that observational campaigns in the future targeting $\sim 31$ mag/arcsec$^2$ are a promising avenue to map the stellar diffuse component in groups and clusters out to at least half the virial radius.

%%%%%%%%%%%%%%%%
%%%%%%%%%%%%%%%%

\section{Summary of Observations} \label{app:obs}

In this section, we present the observational data utilized for the ICL fraction--halo mass relation. Although efforts were made to gather data from sources employing consistent definitions and methods, some studies included in this compilation used different techniques to identify the ICL. For instance, \cite{Spavone2020} studied the Fornax cluster and utilized multi-component fits\footnote{Using multiple equations to describe BCG + ICL (such as single or double S\'{e}rsic profiles and/or exponential profiles).} to distinguish the ICL component, revealing that approximately 34\% of the total light in this cluster originates from the ICL. Another study by \cite{DaRocha2005, DaRocha2008} employed wavelet techniques\footnote{The wavelet technique consists of deconvolving the signal into wavelet coefficients (since images are a 2D signal, each wavelet coefficient corresponds to a plane), identifying the objects representations in the wavelet space, defining the objects with a “multiscale vision model” \citep{Bijaoui1995} and reconstructing the detected objects \citep{DaRocha2005, Jimenez-Teja2016, Ellien2021}.}  to measure the ICL in six objects from the Hickson Compact Group (HCG) catalog, resulting in a wide range of observed ICL fractions. The ICL fractions varied from no detected ICL in HCG 88 to approximately 33\% for HCG 79 (the fraction representing the ratio of ICL light to the total light of the group in the R band). Some studies have combined observational results with semi-analytical simulations to estimate the ICL fraction, such as the deep survey of the Virgo cluster conducted by \cite{Mihos2017}, which estimated a range of 7-15\% for the ICL fraction.

Furthermore, variations in the choice of observational bands used to infer the ICL fraction can be noticed. Different bands can yield different measurements of the ICL fraction. For example, \cite{Ragusa2021} investigated HCG 86 and found an ICL fraction of 14\% in the r band  and 19\% in the g band, highlighting how the choice of band in observations can influence the measured amount of light attributed to the ICL.

An interesting outcome of these measurements is the dispersion observed in the ICL fraction at a given halo mass. \cite{Iodice2020} compared the ICL fractions of several different objects using the same method and demonstrated how features present in the ICL, outer envelope of the brightest cluster galaxy, and the presence of HI can indicate different evolutionary stages and mass assembly histories for different groups and clusters. 

Additionally, some studies have linked the ICL fraction to the mean morphology of the group or cluster. \cite{Poliakov2021} measured the ICL fraction for multiple HCG objects and found that the mean surface brightness of the intra-group light correlates with the mean morphology of the group, with brighter intra-group light observed in groups with a larger fraction of early-type galaxies. It is important to note that the investigated groups in the TNG50 simulations exhibit various morphologies and formation histories, which may contribute to the wide variation in ICL fractions at a given mass. 

Table~\ref{tab:obs_data} below provides a summary of the observational results compiled from multiple sources.

\begin{table*}
\tiny
\caption{Observational Data.}
\label{tab:obs_data}
\begin{tabular}{lllll}
\hline
\hline
\textbf{Object} & \textbf{ref. for f$_{\rm ICL}$} & \textbf{f$_{\rm ICL}$} & \textbf{M$_{\rm halo}$ [$\rm 10^{\rm 13}$ M$_{\odot}$]} & \textbf{Comments}\\
\hline
\hline

Fornax & \cite{Spavone2020} & 34\% (21.4\% - 53.8\%) & 6.3 & \multirow{3}{*}{\parbox{7cm}{ref. for mass: \cite{Spavone2020, Ragusa2021, MontesReview2022} \\ Observation was conducted in u,g,r,i bands \citep{Spavone2020}}}\\
 & \cite{MontesReview2022} & 34\% (19\% - 49\%) & & \\
 & \cite{Ragusa2021} & 34$\pm$2\% & & \\

\hline
Virgo & \cite{Mihos2017} & 7 -15\% & 31.6 (1) & \multirow{2}{*}{\parbox{7cm}{ref. for mass(1): \cite{Ragusa2021, Spavone2020}\\ ref. for mass(2): \cite{Weinmann2011}\\ Observation was conducted in B and V bands \citep{Mihos2017} \\Using results of simulations \citep{Rudick2009}, \cite{Mihos2017} estimated 5-10\% of the ICL should be the form of coherent streams (the relatively high surface brightness tidal features that represent material most recently stripped from their host galaxies), which help them estimate the whole amount of ICL luminosity.}}\\
 & \cite{Ragusa2021} & 11$\pm$3\% & 14-40 (2) & \\
 & & & & \\
 & & & & \\
 & & & & \\
 & & & & \\
 & & & & \\ 

\hline
IC 1459 & \cite{Iodice2020, Ragusa2021} & 2$\pm$2\% & 3.7 & \multirow{2}{*}{\parbox{7cm}{ref. for mass: \cite{Iodice2020, Ragusa2021}\\ Observation was conducted in g and r bands \citep{Iodice2020}}} \\
& & & & \\

\hline
NGC 5018 & \cite{Spavone2018} & 41\% & 0.68 & \multirow{3}{*}{\parbox{7cm}{ref. for mass: \cite{Ragusa2021, Iodice2020, MontesReview2022}\\ Observation was conducted in u,g,r bands \citep{Spavone2018}}} \\
 & \cite{Ragusa2021} & 40$\pm$5\% & & \\
 & \cite{Iodice2020} & 41$\pm$5\% & & \\

\hline
NGC 1533 & \cite{Ragusa2021, Iodice2020} & 8$\pm$2\% & 0.49 & \multirow{2}{*}{\parbox{7cm}{ref. for mass: \cite{Ragusa2021, Iodice2020}\\ Observation was conducted in g,r bands \citep{Cattapan2019}}} \\
 & & & & \\

\hline
HCG 8 & \cite{Poliakov2021, Ragusa2021} & 25.1\% & 5.29 & \multirow{2}{*}{\parbox{7cm}{ref. for mass: \cite{MontesReview2022}\\ Observation was conducted in r band \citep{Poliakov2021}}} \\
 & & & & \\

\hline
HCG 15 & \cite{DaRocha2008, Iodice2020} & \multirow{2}{*}{\parbox{2.5cm}{B: 16$\pm$3\% $\rm ^{\rm a,b}$\\ R: 18$\pm$4\%$^{\rm a,b}$}} & 5.3$\rm ^{b}$ (1) & \multirow{2}{*}{\parbox{7cm}{ref. for mass(1): \cite{DaRocha2008, MontesReview2022}\\ ref. for mass(2): \cite{DaRocha2008, Ragusa2021,  Iodice2020} \\ Observation was conducted in B and R bands \citep{DaRocha2008}}}\\
 & & & & \\
 & & & & \\ 
 & \cite{DaRocha2008, Ragusa2021} & \multirow{2}{*}{\parbox{2.5cm}{B: 19$\pm$4\% $^{\rm a,c}$\\ R: 21$\pm$4\%$^{\rm a,c}$}} & 5.67$\rm ^{c}$ (2) & \\
 & & & & \\

\hline
HCG 17 & \cite{Poliakov2021} & 16.3\% & & Observation was conducted in r band \citep{Poliakov2021} \\

\hline
HCG 35 & \multirow{2}{*}{\parbox{4cm}{\cite{DaRocha2008, Ragusa2021, Iodice2020}}} & \multirow{2}{*}{\parbox{2.5cm}{B: 15$\pm$3\% $^{\rm a}$\\ R: 11$\pm$2\%$^{\rm a}$}} & 1.51 (1) & \multirow{3}{*}{\parbox{7cm}{ref. for mass(1): \cite{DaRocha2008, Ragusa2021, Iodice2020, MontesReview2022}\\ ref. for mass (2): \cite{MontesReview2022} \\ Observation was conducted in B and R bands \citep{DaRocha2008}}}\\
 & & & & \\ \\
 & \cite{Poliakov2021} & 12.8\% & 3.1 (2) & \\

\hline
HCG 37 & \cite{Poliakov2021, Ragusa2021} & 12.7\% & 5.87 (1) & \multirow{3}{*}{\parbox{7cm}{ref. for mass(1): \cite{MontesReview2022} \\ref. for mass(2): \cite{Ragusa2021}\\Observation was conducted in r band \citep{Poliakov2021}}} \\
 & & & 2.24 (2) & \\ \\
% & & & & & \\

\hline
HCG 51 & \cite{DaRocha2008, Iodice2020} & \multirow{2}{*}{\parbox{2.5cm}{B: 26$\pm$5\% $^{\rm a,b}$\\ R: 24$\pm$5\%$^{\rm a,b}$}} & 3.59$^{\rm b}$ (1) & \multirow{2}{*}{\parbox{7cm}{ref. for mass(1): \cite{DaRocha2008, MontesReview2022}\\ ref. for mass(2): \cite{DaRocha2008} \\ Observation was conducted in B and R bands \citep{DaRocha2008}}}\\
 & & & & \\
 & & & & \\ 
 & \cite{DaRocha2008, Ragusa2021} & \multirow{2}{*}{\parbox{2.5cm}{B: 31$\pm$6\% $^{\rm a,c}$\\ R: 28$\pm$5\%$^{\rm a,c}$}} & 0.74$^{\rm c}$ (2) & \\
 & & & & \\

\hline
HCG 74 & \cite{Poliakov2021, Ragusa2021} & 7.5\% & 28.65 & \multirow{2}{*}{\parbox{7cm}{ref. for mass: \cite{MontesReview2022}\\ Observation was conducted in r band \citep{Poliakov2021}}} \\
 & & & & \\

\hline
HCG 79 & \multirow{2}{*}{\parbox{4cm}{\cite{DaRocha2005, Ragusa2021}}} & \multirow{2}{*}{\parbox{2.5cm}{B: 46$\pm$11\% $^{\rm a}$\\ R: 33$\pm$11\%$^{\rm a}$}} & 1.04 (1) & \multirow{3}{*}{\parbox{7cm}{ref. for mass(1): \cite{MontesReview2022} \\ref. for mass(2): \cite{Ragusa2021, Iodice2020}\\Observation was conducted in B and R bands \citep{DaRocha2005}}} \\
& & & & \\ \\
 & \cite{Iodice2020} & 46$\pm$10\% & 3.98 (2) & \\

\hline
HCG 86 & \cite{Ragusa2021} & \multirow{2}{*}{\parbox{2.5cm}{$\rm Radius < 160 kpc:$ \\ g: 19$\pm$3\% (35$\pm$5\% $^{\rm d}$)\\ r: 14$\pm$2\% (29$\pm$6\% $^{\rm d}$)\\ $\rm Radius < 120 kpc:$ \\ g:16$\pm$3\% (28$\pm$5\% $^{\rm d}$)\\ r: 11$\pm$2\% (23$\pm$7\% $^{\rm d}$)}} & 0.85 & \multirow{2}{*}{\parbox{7cm}{ref. for mass: \cite{Ragusa2021}\\ Observation was conducted in g,r,i bands \citep{Ragusa2021}}} \\
 & & & & \\
 & & & & \\
 & & & & \\
 & & & & \\
 & & & & \\
 & & & & \\
 
\hline
HCG 88 & \multirow{2}{*}{\parbox{4cm}{\cite{DaRocha2005, Iodice2020}}} & 0$^{\rm a}$ & 2.88 (1) &  \multirow{3}{*}{\parbox{7cm}{ref. for mass(1): \cite{MontesReview2022}\\ ref. for mass(2): \cite{Iodice2020}\\ Observation was conducted in B and R bands \citep{DaRocha2005}}} \\
 & & & 0.12 (2) & \\
 & & & & \\

\hline
HCG 90 & \cite{Ragusa2021, Iodice2020} & 38$\pm$3\% & 1.17 & ref. for mass: \cite{Ragusa2021, Iodice2020} \\

\hline
HCG 95 & \multirow{2}{*}{\parbox{4cm}{\cite{DaRocha2005, Ragusa2021}}} & \multirow{2}{*}{\parbox{2.5cm}{B: 11$\pm$26\% $^{\rm a}$\\ R: 12$\pm$10\%$^{\rm a}$}} & 2.14 & \multirow{2}{*}{\parbox{7cm}{ref. for mass: \cite{Ragusa2021, MontesReview2022}\\ Observation was conducted in B and R bands \citep{Ragusa2021} \citep{Poliakov2021}}} \\
 & & & & \\

\hline
\multicolumn{3}{l}{$^{\rm a}$ ICL fractions are measured using wavelet technique} \\ 
\multicolumn{3}{l}{$^{\rm b}$ corresponding to sextet configuration} \\
\multicolumn{3}{l}{$^{\rm c}$ corresponding to quintet configuration} \\
\multicolumn{3}{l}{$^{\rm d}$ Measured ICL fraction with respect to mass of the BCG} \\

\end{tabular}
\end{table*}

%%%%%%%%%%%%%%%%%%%%%%%%%%%%%%%%%%%%%%%%%%%%%%%%%%

% Don't change these lines
\bsp	% typesetting comment
\label{lastpage}
\end{document}